\documentclass[journal]{IEEEtran}

\usepackage[utf8]{inputenc}

\usepackage{cite}
\usepackage{multicol}
\usepackage[bookmarks=true]{hyperref}

\usepackage{mathrsfs} 
\usepackage{amsmath,amssymb,amsthm,amsfonts,graphicx,float}
\usepackage{mathtools}
\usepackage{color} 
\usepackage{tabularx}

\usepackage{algorithm}
\usepackage{algpseudocode}

\usepackage{tikz,pgfplots}
\pgfplotsset{compat=1.8}
\usepackage{pgfplotstable}
\usepgfplotslibrary{groupplots}
\usetikzlibrary{spy}
\usepackage{times}

\newcommand{\myqedblock}{\hfill$\square$}
\newcommand{\myet}[1]{#1}
\newtheorem{theorem}{Theorem}
\newtheorem{assumption}{Assumption}
\newtheorem{sideinfo}{Side information}

\newtheorem{corollary}{Corollary}

\newtheorem{definition}{Definition}
\newtheorem{example}{Example}

\newtheorem{lemma}{Lemma}

\newtheorem{problem}{Problem}

\newtheorem{remark}{Remark}

\newcommand{\reachalgo}{\texttt{DaTaReach}}
\newcommand{\controlalgo}{\texttt{DaTaControl}}
\newcommand{\optsolver}{\texttt{AdaRES}}

\begin{document}

\title{\LARGE \bf On-The-Fly Control of Unknown Systems: From Side Information to Performance Guarantees through Reachability
    \thanks{This material is based on work partly supported by Air Force Office of Scientific Research (FA9550-19-1-0005) and National Aeronautics and Space Administration (80NSSC19K0209).\newline 
    \indent F. Djeumou, A. Vinod, and U. Topcu are with the Department of Electrical and Computer Engineering, Oden Institute for Computational Engineering and Sciences, and the Department of Aerospace Engineering and Engineering Mechanics at the University of Texas at Austin, Austin, TX, USA. Email: \texttt{\{fdjeumou, utopcu\}@utexas.edu}, \texttt{aby.vinod@gmail.com}.\newline
    \indent E. Goubault and S. Putot are with the LIX, CNRS, Ecole Polytechnique, Institut Polytechnique de Paris, France. Email: \texttt{\{goubault, putot\}@lix.polytechnique.fr}.}}
\author{Franck Djeumou, Abraham P. Vinod, Eric Goubault, Sylvie Putot, and Ufuk Topcu}
\maketitle

\begin{abstract}
    We develop data-driven algorithms for reachability analysis and control of systems with a priori unknown nonlinear dynamics. The resulting algorithms not only are suitable for settings with real-time requirements but also provide provable performance guarantees. To this end, they merge \myet{noisy} data from only a single finite-horizon trajectory and, if available, various forms of side information. Such side information may include knowledge of the regularity of the dynamics, algebraic constraints on the states, monotonicity, or decoupling in the dynamics between the states. Specifically, we develop two algorithms, \reachalgo{} and \controlalgo{}, to over-approximate the reachable set and design control signals for the system on the fly. \reachalgo{} constructs a differential inclusion that contains the unknown dynamics. Then, in a discrete-time setting, it over-approximates the reachable set through interval Taylor-based methods applied to systems with dynamics described as differential inclusions. We provide a bound on the time step size that ensures the correctness and termination of \reachalgo{}. \controlalgo{} enables convex-optimization-based control using the computed over-approximation and the receding-horizon control framework. Besides, \controlalgo{} achieves near-optimal control and is suitable for real-time control of such systems. We establish a bound on its suboptimality and the number of primitive operations it requires to compute control values. Then, we theoretically show that \controlalgo{} achieves tighter suboptimality bounds with an increasing amount of data and richer side information. Finally, experiments on a unicycle, quadrotor, and aircraft systems demonstrate the efficacy of both algorithms over existing approaches.
\end{abstract}


\section{Introduction} \label{sec:intro}

Consider a scenario in which significant and unexpected changes in the dynamics of a system occur. The changes in the dynamics are such that the a priori known model cannot be used, and there is a need to learn the new dynamics on the fly. In such a scenario, the system has access to data from only its current trajectory and needs to retain a certain degree of control. This paper considers the problem of data-driven, on-the-fly control of systems with unknown nonlinear dynamics under severely limited data.

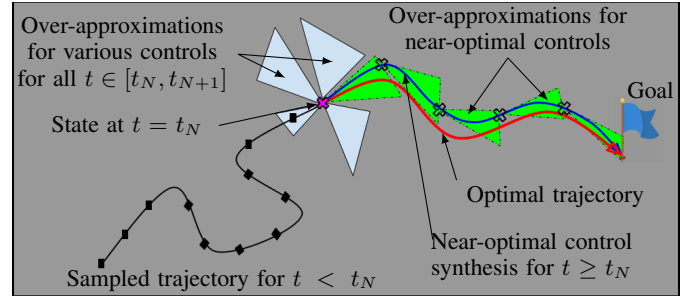
\begin{figure}[t]
    \centering
    \includegraphics[width=1\linewidth]{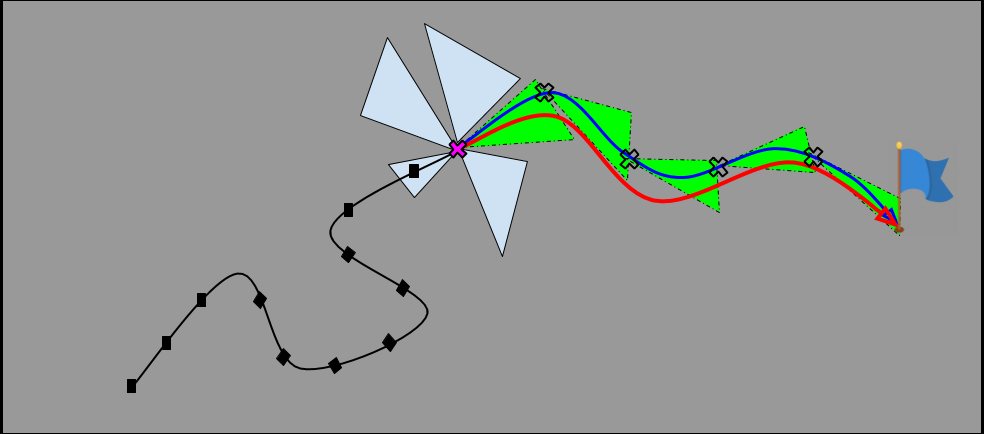} 
    \label{fig:prob_st}
    \begin{tikzpicture}[remember picture, overlay, text
        width=9em]
        \small
        \node [text centered, xshift=13em, yshift=10em]
            (Goal) {Goal};
        \node [text centered, xshift=-9.2em,
            yshift=11.5em] (Overapproximation)
            {Over-approximations for various controls for all $t \in [t_N,t_{N+1}]$};
        \node [text centered, xshift=-9em, yshift=8.5em,
            text width=7em] (System) {State at $t =t_N$};
        \node [text width=18em, text centered,
            xshift=-5em, yshift=2em] (Trajectory)
            {Sampled trajectory for $t< t_N$};
        \node [text centered, xshift=7em, yshift=12.5em, text width=10.5em] 
            (nearoptimal) {Over-approximations for near-optimal controls};
        \node [text centered, xshift=8em, yshift=3em] 
            (Control) {Near-optimal control synthesis for
            $t \geq t_N$};
        \node [text centered, xshift=9em, yshift=5.5em] 
            (Optimal) {Optimal trajectory};
        \draw [-latex]  (System.east) -- 
            ++ (4.15em, 0.9em);
        \draw [-latex]  (Overapproximation.east) -- 
            ++ (2em, -1em);
        \draw [-latex]  (Overapproximation.east) -- 
            ++ (4em, -0.10);
        \draw [-latex]  ([xshift=0em,
            yshift=0.2em]nearoptimal.south) -- ++ (1.7em, -1.8em);
        \draw [-latex]  ([xshift=0em,
            yshift=0.2em]nearoptimal.south) -- ++ (-1.7em, -2.4em);
        \draw [-latex]  ([xshift=0.7em,
            yshift=0.7em]Control.west) -- ++ (-1.2em, 7.0em);
        \draw [-latex]  ([xshift=0.7em,
            yshift=0.7em]Optimal.west) -- ++ (-0.6em, 2em);
    \end{tikzpicture}
    \vspace*{-1em}
    \caption{We use limited data and side information for on-the-fly control of systems with unknown dynamics. At each sampling time, we compute an over-approximation of the set of states the system may reach to describe the uncertainty in its unknown trajectory and construct one-step optimal controllers via convex optimization.}
    \vspace*{-3mm}
\end{figure}

We develop data-driven algorithms, \reachalgo{} and \controlalgo{}, for the reachability analysis and control of systems with a priori unknown dynamics. \reachalgo{} exploits the data from a given and \myet{and ongoing} trajectory of the system to compute an over-approximation of the set of states it may reach. \controlalgo{} incorporates the computed over-approximation into a constrained optimal control problem, which is then solved on the fly.

Specifically, \reachalgo{} and \controlalgo{} can work with \myet{possibly noisy} data from only a single finite-horizon trajectory of the system and take advantage of various forms of side information on the dynamics. More specifically, the data include finite \myet{(noisy)} samples of the states, the derivatives of the states, and the control signals applied. The side information may be a priori knowledge of the regularity of the dynamics, bounds on the vector field locally, monotonicity of the vector field, decoupling in the dynamics among the states, algebraic constraints on the states, or knowledge of parts of the dynamics. For example, such side information may be due to known elementary laws of physics or may be directly extracted from the data, as we illustrate in the numerical simulations.

\reachalgo{} provides closed-form expressions for over-approximations of the reachable set of unknown dynamical systems. It first uses the available data and the given side information to construct differential inclusions that contain the unknown vector field. Unlike existing work~\cite{rezaTAC2020} that provides state-independent differential inclusions, the constructed differential inclusions account for dependencies on the states and control signals. Then, \reachalgo{} provides closed-form expressions for over-approximations of the reachable set associated with these data-driven differential inclusions. Specifically, it builds on interval Taylor-based methods~\cite{berz1998verified,nedialkov1999validated,ericForwardInner2017} to over-approximate the reachable set of dynamics described by the differential inclusions. The closed-form expressions can incorporate the available side information, and increasing amount of data and richer side information provide tighter over-approximations of the reachable set.

The closed-form expressions enable convex-optimization-based, on-the-fly control of unknown dynamical systems through the one-step receding-horizon control framework~\cite{mayne2000constrained}. Specifically, we seek to sequentially minimize a given one-step cost function in a discrete-time setting. We refer to such an optimization problem as a \emph{one-step optimal control} problem. The one-step cost function, which encodes the desired behavior of the system, has to be optimized in a black-box manner since it is typically a function of the unknown future state, in addition to the known current state and the currently applied control input. \controlalgo{} computes \emph{approximate solutions} to the one-step optimal control problem through convex optimization relaxations. The convex relaxations are obtained by replacing the unknown future state with a control-affine linearization of the corresponding over-approximation of the reachable set.

\myet{More specifically, we prove that \controlalgo{} achieves near-real-time and near-optimal control of the unknown system.  Indeed, we provide bounds on the suboptimality of the relaxed convex problems' solutions with respect to the optimal solutions if the dynamics were known. Then, we theoretically show that the obtained bounds on the suboptimality become tighter with an increasing amount of data and richer side information. Besides, we prove that \controlalgo{} is suitable for real-time control of the system through an explicit upper bound on the number of primitive operations required to terminate.}

\myet{Empirically, through a series of experiments on a unicycle, quadrotor, and aircraft systems, we show that \controlalgo{} is significantly more computationally efficient, robust to noise, and less suboptimal than the baselines \texttt{SINDYc}\cite{kaiser2018sparse}, \texttt{DeePC}\cite{Coulson2019DataEnabledPC}, \texttt{C2Opt}\cite{vinod2020convexified}, and \texttt{CGP-LCB}\cite{krause2011contextual}. Indeed, the experiments show that \controlalgo{} is \emph{three to five} orders of magnitude faster than the baselines, achieves the control objective using at least \emph{three times} fewer system's interactions (less suboptimal), and outperforms the baselines in terms of robustness to noise.}

\myet{\noindent {\bf{Contributions.}} 
This paper significantly extends our conference paper~\cite{djeumou2020onthefly} by first augmenting \controlalgo{} and \reachalgo{} with the ability to handle noise in the data. Indeed, the approach in~\cite{djeumou2020onthefly} assumes exact measurements of the state and the state derivative. Then, we provide a bound on the discrete-time step size such that \reachalgo{} and \controlalgo{} are guaranteed to be correct and to terminate. The algorithms in~\cite{djeumou2020onthefly} assume "sufficiently small" time step to work, without any explicit bound to characterize such small time step. Finally, we prove an explicit bound on the number of primitive operations required by \controlalgo{} to terminate and provide additional numerical experiments. With respect to state-of-the-art data-driven techniques, this paper proposes near-optimal and real-time data-driven algorithms that work with an extremely scarce amount of data and that can incorporate side information on the underlying dynamics when learning to control nonlinear systems. We provide performance guarantees such as bounds on the suboptimality and computational complexity of the algorithms. At last, through extensive comparisons with state-of-the-art baselines, we empirically demonstrate the efficiency of the proposed approach.}

\noindent {\textbf{Related work.}
\myet{Several approaches for data-driven control} combine system identification with model predictive control~\cite{korda2018linear,kaiser2018sparse,proctor2016dynamic,ornik2019myopic,vinod2020convexified}. In~\cite{korda2018linear}, the authors use Koopman theory to lift the unknown nonlinear dynamics to a higher-dimensional space where they perform linear system identification. \texttt{SINDYc}~\cite{kaiser2018sparse} utilizes a sparse regression over a library of nonlinear functions for nonlinear system identification. \texttt{DMDc}~\cite{proctor2016dynamic} uses the spectral properties of the collected data to obtain approximate linear models. Myopic control~\cite{ornik2019myopic} uses a finite sequence of perturbations to learn a local linear model, which is then used to optimize a model-driven goodness function that encodes desirable behaviors. \texttt{DMDc}, \texttt{SINDYc}, \myet{Myopic control}, and approaches based on the Koopman theory require significantly more data than the proposed approach\myet{, do not provide real-time guarantees, and, in their current forms, cannot incorporate any of the side information in this paper.} 


Contextual optimization-based approaches tackle the data-driven control problem via surrogate optimization and skip the system identification step~\cite{vinod2020convexified,krause2011contextual}. These approaches iteratively minimize the one-step cost function in a black-box manner using the data. \texttt{C2Opt}~\cite{vinod2020convexified} exploits the structure in the given problem and utilizes side information (for example, smoothness) and convex optimization to solve this problem. It overcomes the drawbacks of the traditional Gaussian process-based approaches~\cite{krause2011contextual}, namely high computational costs, expensive hyperparameter tuning, and inability to incorporate side information. Unlike \controlalgo{}, \texttt{C2Opt} considers limited forms of side information and relies on the knowledge of the gradient of the one-step cost, that may not be accessible. 

\myet{Recent work~\cite{Coulson2019DataEnabledPC,berberich2020data,berberich2020combining,markovsky2021data,van2020noisy,van2020data} have proposed data-driven control techniques for unknown dynamical systems that bypass the system identification step. These techniques mostly assume linear time-invariant dynamical systems and are extremely performant in such a setting. In contrast, the approach in this paper works with general unknown nonlinear dynamical systems and can incorporate side information on the underlying dynamics. Besides, on a linear version of the aircraft dynamics, we provide a comparison with \texttt{DeePC}~\cite{Coulson2019DataEnabledPC}, which is a traditional baseline using the same behavioral systems theory foundation~\cite{1428856} as the approaches in~\cite{berberich2020data,berberich2020combining,markovsky2021data,van2020noisy,van2020data}. We show that \controlalgo{} is \emph{two orders} of magnitude faster than \texttt{DeePC} while being more accurate with a low amount of data and as accurate with a high amount of data.}

The authors of~\cite{devonport2020data,haesaert2017data,chakrabarty2018data} have considered the problem of data-driven estimation of the reachable sets of partially unknown dynamical systems. The approaches in these work rely on system identification using either supervised learning algorithms~\cite{chakrabarty2018data} or Gaussian process-based algorithms~\cite{haesaert2017data, devonport2020data}. Such approaches are unable to take advantage of the side information, require significantly more data than \reachalgo{}, and provide only probabilistic guarantees of the correctness of the computed reachable sets. \myet{Instead, \reachalgo{} provides correct over-approximations of the reachable set at the expense of sometimes being conservative.} }

\section{Preliminaries} \label{sec:prelem}
We denote an interval by $[a , b] = \{ x \in \mathbb{R} |  a \leq x \leq b \}$ for some $a, b\in \mathbb{R}$ such that $a \leq b$, the set $\{i,\hdots,j\}$ by $\mathbb{N}_{[i,j]}$ for $i,j \in \mathbb{N}$ with $i\leq j$ , the $2$-norm by $||\cdot||_2$, the $k^\mathrm{th}$ component of a vector $x$ and the $(k,j)$ component of a matrix $X$ by ${(x)}_k$ and ${(X)}_{k,j}$, respectively, and the Lipschitz constant of $f : \mathcal{X} \to \mathbb{R}$ by $L_f = \sup \{ L \in \mathbb{R} \: \vert \: |f(x) - f(y)| \leq L \|x-y\|_2, x,y \in \mathcal{X}, x \neq y\}$ for $\mathcal{X} \subseteq \mathbb{R}^n$. A function $f \in \mathscr{C}^k(\mathcal{X})$, referred as $f$ is $\mathcal{C}^{k}$, with $k\geq 0$ if $f$ is continuous on $\mathcal{X} \subseteq \mathbb{R}^n$ and all the partial derivatives of order $1,\hdots,k$ exist and are continuous on $\mathcal{X}$, and $f$ is \emph{piecewise-}$\mathscr{C}^k$ with $k \geq 0$ if there exists a partition of $\mathcal{X}$ such that $f$ is $\mathscr{C}^k$ on each set in the partition. Comparisons (e.g., $\ge$) between matrices or vectors are conducted elementwise.

\subsection{Interval Analysis}\label{sec:prelem-interval-analysis}

We denote the set of intervals on $\mathbb{R}$ by $\mathbb{IR} = \{ \mathcal{A} = [\underline{\mathcal{A}},\overline{\mathcal{A}}]  \: | \: \underline{\mathcal{A}},\overline{\mathcal{A}} \in \mathbb{R}, \underline{\mathcal{A}} \leq \overline{\mathcal{A}}\}$, the set of $n$-dimensional interval vectors by $\mathbb{IR}^n$, and the set of $n\times m$-dimensional interval matrices by $\mathbb{IR}^{n \times m}$. We carry forward the definitions~\cite{moore1966interval} of arithmetic operations, set inclusion, and intersections of intervals to interval vectors and matrices by applying them componentwise. We denote the absolute value of an interval $\mathcal{A}$ by $|\mathcal{A}| = \max \{ |\overline{\mathcal{A}}|, |\underline{\mathcal{A}}| \}$, the infinity norm of $\mathcal{A} \in \mathbb{IR}^n$ by $\|\mathcal{A}\|_{\infty} = \sup_{i \in \mathbb{N}_{[1,n]}} |(\mathcal{A})_i|$, the Cartesian product between intervals by $\otimes$, i.e., $ \mathcal{A}\otimes \mathcal{B}={[[\underline{\mathcal{A}},\overline{\mathcal{A}}], [\underline{\mathcal{B}},\overline{\mathcal{B}}]]}\in \mathbb{IR}^2$ for any two intervals $ \mathcal{A}, \mathcal{B}\in \mathbb{IR}$. We define $ \mathcal{A}^n$ as the Cartesian product of any interval $ \mathcal{A}\in \mathbb{IR}$ with itself $n$ times. We use the term interval to specify an interval vector or interval matrix when it is clear from the context.

Given $f : \mathcal{X} \mapsto \mathcal{Y}$ with $\mathcal{X} \subseteq \mathbb{R}^n$ and $\mathcal{Y} \subseteq \mathbb{R}^{m}$ (or $\mathcal{Y} \subseteq \mathbb{R}^{n \times m}$), we define an \emph{interval extension of $f$} as \myet{\emph{an} interval-valued function $\boldsymbol{f} : \mathbb{IR}^n \mapsto \mathbb{IR}^n$ such that}
\begin{align}
     \boldsymbol{f}(\mathcal{A}) \supseteq \mathscr{R}(f, \mathcal{A}) = \{ f(x) \: | \: x \in \mathcal{A} \} , \quad \forall \mathcal{A} \subseteq \mathcal{X}. \label{eq:range-def-over-approximation}
\end{align}
Thus, given an interval $ \mathcal{A}$, $\boldsymbol{f}(\mathcal{A})$ is an interval that over-approximates the range of values taken by $f$ over $\mathcal{A}$. We denote interval-valued vector functions via bold lowercase and interval-valued matrix functions via bold uppercase symbols.
\begin{example}[\textsc{Interval extension of $2-$norm}] \label{ex:norm-2-extension}
    Consider $f=||\cdot||_2$. We compute its interval extension $\boldsymbol{f}$ via interval extensions of $\alpha=\sqrt{\cdot}$ and $\beta={(\cdot)}^2$. For any $ \mathcal{A}=[\underline{\mathcal{A}}, \overline{\mathcal{A}}]\in \mathbb{IR}$,
    \begin{align}
        \boldsymbol{\alpha}(\mathcal{A}) &= [\sqrt{\underline{\mathcal{A}}}, \sqrt{\overline{\mathcal{A}}}], \: \: \: \text{if } \underline{\mathcal{A}} \geq 0 , \label{eq:sqrt_ext}\\
        \boldsymbol{\beta}(\mathcal{A}) &= \begin{cases} [0 , \max\{\underline{\mathcal{A}}^2,\overline{\mathcal{A}}^2\}],& \text{if } 0 \in \mathcal{A} \\ 
        [\min \{\underline{\mathcal{A}}^2,\overline{\mathcal{A}}^2\}, \max\{\underline{\mathcal{A}}^2,\overline{\mathcal{A}}^2\}],& \text{otherwise}.\end{cases}\label{eq:sqr_ext}
    \end{align}
    Using \eqref{eq:sqrt_ext} and \eqref{eq:sqr_ext} with interval arithmetic, we have, for any $ \mathcal{S}=[\mathcal{S}_1,\hdots, \mathcal{S}_n] \in \mathbb{IR}^n$, $\boldsymbol{f}( \mathcal{S}) = \boldsymbol{\alpha}\left({\sum_{i=1}^n \boldsymbol{\beta}( \mathcal{S}_i)}\right)$.
\end{example}

\subsection{Over-Approximations of the Reachable Set} \label{sec:prelem-taylor-method}

Consider a nonlinear dynamical system,
\begin{equation} \label{general_dynamic}
    \dot{x} = h (x , u),
\end{equation}
where the state $x : \mathbb{R}_+ \mapsto \mathcal{X}$ is a continuous-time signal evolving in $\mathcal{X} \in \mathbb{IR}^n$, the control $u \in \mathbb{U}$ is a signal of time evolving in the control set $\mathcal{U}\in\mathbb{IR}^m$ with $\mathbb{U}=\{v : \mathbb{R}_+ \mapsto \mathcal{U}  \: | \: v \text{ is \emph{piecewise-}}\mathscr{C}^{D_u} \}$ for $D_u \geq 0$, and the function $h : \mathcal{X} \times \mathcal{U} \mapsto \mathcal{Y}$ is $\mathscr{C}^{D_h}$ for some $D_h\geq 1$ and $\mathcal{Y} \subseteq \mathbb{R}^n$.

\begin{definition}[\textsc{Trajectory of the System}]
    Given an initial state $x_i = x(t_i)$ at time $t_i$ and a control signal $u \in \mathbb{U}$, a trajectory of~\eqref{general_dynamic} is a continuous function of time
    $x(\cdot;x_i, u) : [t_i,\infty[ \mapsto \mathcal{X}$ that satisfies~\eqref{general_dynamic}.
\end{definition}

We are interested in the set of states reachable by trajectories of the system when the initial state and the control signal are uncertain quantities. We call such a set the reachable set of the system, and we define it as follows. 
\begin{definition} [\textsc{Reachable set}]
    Given a set $\mathcal{I}_i \subseteq \mathcal{X}$ of states  at time $t_i$ and a set $\mathbb{V} \subseteq \mathbb{U}$ of control signals, the reachable set of the dynamics~\eqref{general_dynamic} at time $t\geq t_i$ is given by $\mathcal{R}(t, \mathcal{I}_i,\mathbb{V}) =\{ z \in  \mathcal{X} \: \vert \:  \exists x_i \in \mathcal{I}_i, \: \exists v \in \mathbb{V}, \: z=x(t;x_i,v)\}.$
\end{definition}

Given a set $\mathbb{V} \subseteq \mathbb{U}$ and a set $\mathcal{I}_0 \subseteq \mathcal{X}$ of states at time $t_0$, we compute over-approximations of $\mathcal{R}(t, \mathcal{I}_0,\mathbb{V})$ at time $t \geq t_0$ using interval Taylor-based methods~\cite{moore1966interval,nedialkov1999validated,goubault2019inner}. Specifically, we consider a time grid $t_0 <\cdots <t_N$ such that for all $v \in \mathbb{V}$, $v \text{ is } \mathscr{C}^{D_u}$ on each interval $[t_i, t_{i+1}[$. We want to compute sets $\mathcal{R}_i^+\in \mathbb{IR}^n$ (with $\mathcal{R}_0^+=\mathcal{I}_0$) such that for all $i \in \mathbb{N}_{[0,N-1]}$, $\myet{\mathcal{R}_{i+1} = \mathcal{R}(t_{i+1}, \mathcal{R}_{i},\mathbb{V}) \subseteq \mathcal{R}_{i+1}^+}$.
First, interval arithmetic enables to inductively define $\boldsymbol{h}^{[d]}$, the interval extensions of the Taylor coefficients $h^{[d]}$ given by
\begin{align}
    h^{[1]} = h, \: h^{[d+1]} = \frac{1}{d+1}\Big( \frac{\partial h^{[d]}}{\partial x} h + \sum_{l=0}^{d-1} \frac{\partial h^{[d]}}{\partial u ^{(l)}} u^{(l+1)} \Big), \label{eq:taylor-coeff}
\end{align}
\myet{where $u^{(l)}$ is the $l^{\mathrm{th}}$ derivative of $u$ and $h^{[d]}$ is a function of higher order derivatives of $h$.} Next, we start with $\mathcal{R}_0^+= \mathcal{I}_0$, and compute the over-approximations $\{\mathcal{R}_{i}^+\}_{i=1}^N$ iteratively as
\begin{align}
    \mathcal{R}_{i+1}^+= \mathcal{R}_{i}^+ & + \sum\nolimits_{d=1}^{D-1} \big( t_{i+1}-t_{i}\big)^d \big( \boldsymbol{h}^{[d]}(\mathcal{R}_{i}^+, \boldsymbol{v})\big) (t_i) \nonumber \\
                     &+ \big(t_{i+1}-t_{i}\big)^{D} \big( \boldsymbol{h}^{[D]}(\mathcal{S}_{i},\boldsymbol{v})\big)([t_i,t_{i+1}]), \label{eq:taylor-expansion}
\end{align}
where $D\leq\min(D_u+1,D_h)$ is the order of the Taylor expansion, we denote $\boldsymbol{v}^{(0)}(\mathcal{A})$ by $\boldsymbol{v}(\mathcal{A})$ and the intervals $\boldsymbol{v}^{(d)}(\mathcal{A})$ for all $d\in \mathbb{N}_{[0,D_u]}$ are such that $\cup_{v \in \mathbb{V}} \mathscr{R}(v^{(d)},\mathcal{A}) \subseteq \boldsymbol{v}^{(d)}(\mathcal{A})$ with $\mathcal{A} \subseteq \mathbb{R}_+$. In other words, the interval $\boldsymbol{v}^{(d)}(\mathcal{A})$ over-approximates the range of the $d^{\mathrm{th}}$  derivative of all functions $v \in \mathbb{V}$ on the interval $\mathcal{A}$. Note that $\boldsymbol{v}^{(d)}(\mathcal{A})$ is used in the computation of $\boldsymbol{h}^{[p]}$~\eqref{eq:taylor-coeff} for all $p \in \mathbb{N}_{[d+1,D]}$ with $\mathcal{A} = t_i$  and $ \mathcal{A} =  [t_i,t_{i+1}]$. Here, the set $\mathcal{S}_{i} \subseteq \mathcal{X}$ is an \emph{a priori rough enclosure} of $\mathcal{R}(\cdot, \mathcal{R}_i^+, \mathbb{V})$ on $[t_i, t_{i+1}]$ and is a solution of
\begin{align}
    \mathcal{R}_{i}^+ \ + \  [0, t_{i+1}-t_{i}]\ \mathscr{R}(h, \mathcal{S}_{i} \times \boldsymbol{v}([t_i,t_{i+1}])) \ \subseteq \ \mathcal{S}_{i}. \label{eq:rough-enclosure}
\end{align}

Since it is a hard problem to compute exactly the range $\mathscr{R}(h, \mathcal{S}_{i} \times \boldsymbol{v}([t_i,t_{i+1}])$, we instead compute the a priori rough enclosure $\mathcal{S}_i$ recursively by solving
\begin{align}
    \mathcal{R}_{i}^+ \ + \  [0, t_{i+1}-t_{i}]\ \boldsymbol{h}(\mathcal{S}_{i},\boldsymbol{v}([t_i,t_{i+1}])) \ \subseteq \ \mathcal{S}_{i}, \label{eq:rough-enclosure-approx}
\end{align}
where $\boldsymbol{h}$ is an interval extension of the known function $h$. Note that a solution $\mathcal{S}_i$ of~\eqref{eq:rough-enclosure-approx} is also a solution of~\eqref{eq:rough-enclosure} by definition of the interval extension $\boldsymbol{h}$. The existence of the a priori rough enclosure $\mathcal{S}_i$ is guaranteed for a sufficiently small time step size $t_{i+1}-t_{i}$. Hence, picking adequately the time step size is primordial in order to find $\mathcal{S}_i$ in a finite number of iterations when solving the fixed-point equation~\eqref{eq:rough-enclosure-approx}.

\subsection{One-Step Optimal Control}\label{prelem-one-step-control}
Constrained receding-horizon control~\cite{mayne2000constrained} is a form of constrained control in which, at each sampling time, we obtain the current control input by solving a finite-horizon open-loop optimal control problem, using the current state of the system as the initial state. Consider a system with dynamics in the form~\eqref{general_dynamic}, the desired behavior of the system is generally encoded as a \emph{one-step cost function} $c : \mathcal{X} \times \mathcal{U} \times \mathcal{X} \to \mathbb{R}$ that assigns preferences over the tuple of the current state, the current control input, and the corresponding state at the next time step.  Given a constant time step size $\Delta t \geq 0$, we define the next state $x_{p+1} = x(t_{p+1}; x_p, u_p)$ as the value of the trajectory of the system at time $t_{p+1} = t_p + \Delta t$, given the initial state $x_p \in \mathcal{X}$ at time $t_p$ and the constant control signal $u_p \in \mathcal{U}$. Therefore, the finite horizon open-loop optimal control problem at a state $x_i \in \mathcal{X}$ at time $t_i$ is given by 
\begin{align}
        \underset{u_i,\hdots,u_{i+N-1}\in \mathcal{U}}{\mathrm{minimize}}& \quad \sum\nolimits_{p=i}^{i+N-1} c(x_p, u_p, x_{p+1}),  \label{eq:n-step-optimal-control}
\end{align}
where $N \geq 1$ is the planning horizon of the control problem. Note that even if the cost function $c$ is convex, the optimization problem~\eqref{eq:n-step-optimal-control} is generally nonconvex since the next state $x_{p+1}$ is possibly highly nonlinear in the control variable $u_p$ due to the nonlinear dynamics function $h$.

In the setting where the dynamics of the system are unknown, we cannot compute its trajectories, and $c(x_p, u_p, x_{p+1})$ is therefore unknown. Hence, the optimization problem~\eqref{eq:n-step-optimal-control} is similar to the problem of optimizing an a priori unknown function. In such a setting, the inability to accurately predict the next states of the system motivates using the short planning horizon $N=1$. Thus, at each time $t_i$, we solve the \emph{one-step optimal control} problem~\cite{kocijan2004gaussian,vinod2020convexified,park1999robust} given by
\begin{align}
    \underset{u_i \in \mathcal{U}}{\mathrm{minimize}}& \quad c(x_i, u_i, x_{i+1}),
    \label{eq:one-step-optimal-control}
\end{align}
where $x_i$ is the state of the system at $t_i$, $x_{i+1}$ is the unknown next state at time $t_{i+1} = t_i + \Delta t$, and $\Delta t$ is the constant time step size. The reduction to a one-step optimal control problem greatly improves the tractability of the problem of controlling the unknown dynamical system~\cite{kocijan2004gaussian}.  

\section{Problem Statement} \label{sec:problem-statement}

In this paper, we consider \emph{control-affine} nonlinear systems,
\begin{align}
    \dot{x} = f(x) + G(x) u, \label{eq:control-linearization}
\end{align}
where $f : \mathcal{X} \mapsto \mathbb{R}^n$ and $G : \mathcal{X} \mapsto \mathbb{R}^{n \times m}$ are \emph{unknown} vector-valued and matrix-valued functions, respectively, for $\mathcal{X} \in \mathbb{IR}^n$. Note that even though we consider control-affine dynamics, in the general case, we can construct a control-affine model of the system locally and apply the results of the paper.

\begin{assumption}[\textsc{Lipschitz system}]\label{assum:smooth}
    $f$ and $G$  are locally Lipschitz-continuous functions. That is, each component of $f$ and $G$ has a finite Lipschitz constant on each subset of $\mathbb{R}^n$.
\end{assumption}
Assumption~\ref{assum:smooth} is common in the frameworks of reachability analysis and receding-horizon control. Since the domain $\mathcal{X} \in \mathbb{IR}^n$ is bounded, $f$ and $G$ are globally Lipschitz-continuous on $\mathcal{X}$. We exploit such global Lipschitz continuity on $\mathcal{X}$ by \myet{assuming known upper bounds on the Lipschitz constants. That is, we have access to $L_f \in  \mathbb{R}_+^n$ and $L_G \in \mathbb{R}_+^{n \times m}$ with $(L_f)_k = L_{f_k}$ and $(L_G)_{k,l} = L_{G_{k,l}}$ as known upper bounds on the Lipschitz constants of $(f)_k$ and $(G)_{k,l}$ for all $k \in \mathbb{N}_{[1,n]}$ and $l \in \mathbb{N}_{[1,m]}$. We emphasize that the Lipschitz bounds can be directly estimated from data at the expense of weakening some of the guarantees in this paper.}

Let $N \in \mathbb{N}$, $N\geq 1$. Let $\myet{\mathscr{T}_N = \{(\Tilde{x}_i,\Tilde{\dot{x}}_i, u_i)\}_{i=1}^{N}}$ denote a single finite-horizon trajectory containing $N$ \myet{noisy} samples of the exact state $x_i=x(t_i)$, the derivative $\dot{x}_i=\dot{x}(t_i)$ of the state, and the control signal $u_i = u(t_i)$ from a trajectory of the system. \myet{We build on the widely-used \emph{bounded} noise assumption and consider that the following bounds hold:
\begin{align}
    |x(t) - \Tilde{x}(t)| \leq \eta, \: |\dot{x}(t) - \Tilde{\dot{x}}(t)| \leq \bar{\eta},\label{eq:bounded-noise}
\end{align}
for all $t \in \mathbb{R}_+$ and for some vector values $\eta, \bar{\eta} \in \mathbb{R}_+^n$. Here the absolute value and the comparison are conducted elementwise.}

\myet{Given the current noisy measurements of the state $\Tilde{x}_{N+1} = \Tilde{x}(t_{N+1})$ and the trajectory $\mathscr{T}_N$, we first seek to over-approximate the reachable set of the system.}
\begin{problem}[\textsc{Reachable set over-approximation}] \label{prob:reachability}
    Given a set $\mathbb{V} \subseteq \mathbb{U}$ of admissible control signals, a time step size $\Delta t > 0$, and a maximum number  $T > N$ of time steps, compute an over-approximation of the reachable set at time $t_i = t_N + (i-N) \Delta t$ for all $ i \in \mathbb{N}_{[N+1,T]}$.
\end{problem}


Next, we seek to control the system by computing approximate solutions to the one-step optimal control problem~\eqref{eq:one-step-optimal-control}.
\begin{problem}[\textsc{Approximate one-step optimal control}] \label{prob:control}
     Given a single finite-horizon trajectory $\mathscr{T}_{i-1}$ for some $i\in \mathbb{N}, \: i > N$, the current measurement $\Tilde{x}_i$ at time $t_i$, compute, at sampling time $t_i$, approximate solutions of the one-step optimal control problem~\eqref{eq:one-step-optimal-control} and characterize the suboptimality of such solutions with respect to the solutions of the optimal control problem if the dynamics were known.
\end{problem}
\begin{example}[\textsc{Unicycle System}] \label{ex:unicycle-system}
    Consider, as a running example, the unicycle system with dynamics given by
    \begin{equation} \label{eq:unicycle-dynamic}
        \dot{p}_x = v \cos (\theta), \ \dot{p}_y = v \sin (\theta),\ \dot{\theta} = \omega,
    \end{equation}
    where the components of the state $x = [p_x, p_y, \theta]$ represent, respectively, the position in the $x$ plane, the $y$ plane, and the heading of the unicycle. The components of the control $u = [v , \omega]$ represent the speed and the turning rate, respectively. We consider the constraint set $\mathcal{U} = [-3,3]\times [-\pi,\pi]$. In the control-affine form~\eqref{eq:control-linearization}, we have $f = 0$, $(G)_{1,1} = \cos (\theta)$, $(G)_{2,1} = \sin (\theta)$, $(G)_{3,2} = 1$, and $(G)_{k,l} = 0$ otherwise. We assume the dynamics~\eqref{eq:unicycle-dynamic} are unknown and are given the loose Lipschitz bounds $L_f = [0.01,0.01,0.01]$, $L_{G_{1,1}} = L_{G_{2,1}} = 1.1, L_{G_{3,2}} = 0.1$, and $L_{G_{k,l}} = 0$ otherwise. These bounds encode the knowledge about the influence of the current state and control signals on the unicycle dynamics. For example, since $p_x$ and $p_y$ are unchanged when no velocity is applied, we can deduce that  $(G)_{1,2} = (G)_{2,2} = 0$ and therefore $L_{G_{1,2}}$ and $L_{G_{2,2}}$. The same reasoning allows to say that $f=0$, but we do not use such knowledge. Furthermore, we consider the knowledge that the vector field does not depend on its positions. That is, $f(x) = f(\theta)$ and $G(x) = G(\theta)$.
\end{example}

\section{Over-Approximations of the Reachable Set of Unknown Dynamical Systems}
We develop \reachalgo{} to address Problem~\ref{prob:reachability}. It constructs a differential inclusion that contains the unknown vector field. Then, it utilizes the interval Taylor-based method to over-approximate the reachable set of dynamics described by the constructed differential inclusion. We also provide a bound on the time step size that guarantees that an a priori rough enclosure exists. Henceforth, the bound enables to ensure the correctness and termination of \reachalgo{}.

\subsection{Differential Inclusion based on Lipschitz Continuity}
First, to aid the construction of the differential inclusion, we over-approximate $f$ and $G$ at each data point of $\mathscr{T}_N$.
\begin{lemma}[\textsc{Contraction via data}] \label{lem:contraction}
    \myet{Given a data point $(\Tilde{x}_i,\Tilde{\dot{x}}_i, u_i)$, an interval $\mathcal{F}_i \in \mathbb{IR}^n$ such that $f(\Tilde{x}_i) \in \mathcal{F}_i$, and an interval $\mathcal{G}_i \in \mathbb{IR}^{n \times m}$ such that $G(\Tilde{x}_i) \in \mathcal{G}_i$, the intervals $C_{\mathcal{F}_i}$ and $C_{\mathcal{G}_i}$, defined sequentially for $l=1,\hdots,m$ by
    \begin{equation} \label{eq:contraction-fG}
        \begin{aligned}
            (C_{\mathcal{F}_i})_k &= (\mathcal{F}_i)_k \: \cap \: ([\Tilde{\dot{x}}_i - \bar{\eta},\Tilde{\dot{x}}_i + \bar{\eta}] - \mathcal{G}_i u_i)_k, \\
            (s_0)_k &=  ([\Tilde{\dot{x}}_i - \bar{\eta},\Tilde{\dot{x}}_i + \bar{\eta}] - C_{\mathcal{F}_i})_k \cap (\mathcal{G}_i u_i)_k, \\
            (C_{\mathcal{G}_i})_{k,l} &= \begin{cases} \big( ((s_{l-1})_k - \sum_{p>l} (\mathcal{G}_i)_{k,p} (u_i)_p) \cap  ((\mathcal{G}_i)_{k,l} u^l_i) \big) \frac{1}{u^l_i},\\
                                                \quad \text{if } u^l_i = (u_i)_l \neq 0 \\
                                                (\mathcal{G}_i)_{k,l}, \quad \text{otherwise},
                                    \end{cases}\\
            (s_l)_k &= \big((s_{l-1})_k - (C_{\mathcal{G}_i})_{k,l} (u_i)_l \big) \cap \big( \sum_{p > l} (\mathcal{G}_i)_{k,p} (u_i)_p \big), 
        \end{aligned}
    \end{equation}
    for all $k\in \mathbb{N}_{[1,n]}$, are the smallest intervals enclosing $f(\Tilde{x}_i)$ and $G(\Tilde{x}_i)$, respectively, given only the data point, $\mathcal{F}_i$, and $\mathcal{G}_i$.}
\end{lemma}

\myet{The proof for Lemma~\ref{lem:contraction} is derived exactly as in our previous work~\cite{djeumou2020onthefly} for noiseless data, where we replace the unknown exact derivative $\dot{x}_i$ with the uncertain quantities $\Tilde{\dot{x}}_i + [-\bar{\eta}, \bar{\eta}]$ and propagate the result with interval arithmetic. Intuitively, we use the constraint $\dot{\Tilde{x}}_i = f(\Tilde{x}_i) + G(\Tilde{x}_i) u_i$, where $|\dot{\Tilde{x}}_i-\Tilde{\dot{x}}_i| \leq \bar{\eta}$, to remove from $\mathcal{F}_i$ and $\mathcal{G}_i$ some values of $f(\Tilde{x}_i)$ and $G(\Tilde{x}_i)$ that do not satisfy the constraint. Note that, in practise, $|\dot{\Tilde{x}}_i-\Tilde{\dot{x}}_i| \leq \bar{\eta}$ holds since measurements of the state derivative are usually obtained from derivatives of the noisy measurements of the state $\Tilde{x}_i$.
We also provide an illustration of Lemma~\ref{lem:contraction} in Appendix~\ref{appendix-A} for the unicycle system. }

\myet{Next, we provide functions to globally over-approximate $f$ and $G$, via uncertain knowledge of $f$ and $G$ at the data points.}
\begin{lemma}[\textsc{Over-approximation of $f$ and $G$}]\label{lem:overapprox-f-G}
    \myet{Given a set $\mathscr{E}_N = \{ (\Tilde{x}_i,C_{\mathcal{F}_i}, C_{\mathcal{G}_i})\: | \: f(\Tilde{x}_i) \in C_{\mathcal{F}_i}, G(\Tilde{x}_i) \in C_{\mathcal{G}_i}\}_{i=0}^{N}$ and the bounds $L_f$ and $L_G$, the functions $\boldsymbol{f} : \mathcal{X} \to \mathbb{IR}^{n}$ and $\boldsymbol{G} : \mathcal{X} \to \mathbb{IR}^{n \times m}$, given for all $k\in \mathbb{N}_{[1,n]}$ and $l \in \mathbb{N}_{[1,m]}$ by
    \begin{equation}\label{eq:overapprox-f-G}
    \begin{aligned}
        (\boldsymbol{f}(x))_k &=  \bigcap_{(\Tilde{x}_i, C_{\mathcal{F}_i}, \cdot) \in \mathscr{E}_N} (C_{\mathcal{F}_i})_k + L_{f_k} \| x - \Tilde{x}_i\|_2 [-1,1], \\
        (\boldsymbol{G}(x))_{k,l} &= \bigcap_{(\Tilde{x}_i, \cdot, C_{\mathcal{G}_i}) \in \mathscr{E}_N} (C_{\mathcal{G}_i})_{k,l} + L_{G_{k,l}} \| x - \Tilde{x}_i\|_2 [-1,1],
    \end{aligned}
    \end{equation}
    are such that $f(x) \in \boldsymbol{f}(x)$ and $G(x) \in \boldsymbol{G}(x)$ for all $x \in \mathcal{X}$.}
\end{lemma}

\algdef{SE}[DOWHILE]{Do}{doWhile}{\algorithmicdo}[1]{\algorithmicwhile\ #1}%
\begin{algorithm}[!t]
    \caption{Optimal over-approximation of the values of $f$ and $G$ at each data point of a finite-horizon trajectory.}\label{algo:overapprox-datapoints}
    \begin{algorithmic}[1]    
        \Require{Single trajectory $\mathscr{T}_N$, sufficiently large $M >0$, upper bounds on the Lipschitz constants, \myet{noise bound $\bar{\eta}$}.}
    \Ensure{\myet{${\mathscr{E}_N = \{ (\Tilde{x}_i,C_{\mathcal{F}_i}, C_{\mathcal{G}_i}) |  f(\Tilde{x}_i) \in C_{\mathcal{F}_i}, G(\Tilde{x}_i) \in C_{\mathcal{G}_i}\}_{i=0}^{N}}$}}
        \State \myet{$\mathcal{A} \gets \mathcal{X}$, $\mathcal{R}^{f_\mathcal{A}} \gets [-M,M]^n$, $\mathcal{R}^{G_\mathcal{A}} \gets [-M,M]^{n \times m}$} 
        \State \myet{Define $\Tilde{x}_0 \in \mathcal{A}$, $C_{\mathcal{F}_0} \gets \mathcal{R}^{f_\mathcal{A}}$, and $C_{\mathcal{G}_0} \gets \mathcal{R}^{G_\mathcal{A}}$}
        \For{$i \in \mathbb{N}_{[1,N]} \wedge \myet{(\Tilde{x}_i, \Tilde{\dot{x}}_i, u_i)} \in \mathscr{T}_{N}$} \label{alg:begin-init-e}
                \State \myet{Compute $\mathcal{F}_i = \boldsymbol{f}(\Tilde{x}_i), \mathcal{G}_i = \boldsymbol{G}(\Tilde{x}_i)$ via~\eqref{eq:overapprox-f-G} and $\mathscr{E}_{i-1}$} \label{alg:update-ei}
                \State Compute $C_{\mathcal{F}_i}, C_{\mathcal{G}_i}$ via~\eqref{eq:contraction-fG}, $\mathcal{F}_i, \mathcal{G}_i$, and \myet{$(\Tilde{x}_i,\Tilde{\dot{x}}_i,u_i)$} \label{alg:contraction-f-G}
        \EndFor \label{alg:end-init-e}
        \Do \label{alg:while-begin}
            \State Execute lines~\ref{alg:begin-init-e}--\ref{alg:end-init-e} with $\mathscr{E}_N$ instead of $\mathscr{E}_{i-1}$ on line~\ref{alg:update-ei} 
        \doWhile{$\mathscr{E}_N$ is not invariant} \label{alg:while-end}
    \State \Return $\mathscr{E}_N$
  \end{algorithmic}
\end{algorithm}

\myet{}We provide a proof for Lemma~\ref{lem:overapprox-f-G} in our previous work~\cite{djeumou2020onthefly}, where we use interval arithmetic with the definition of the upper bounds on the Lipschitz constants to over-approximate the unknown functions $f$ and $G$. Note that interval extensions of $\boldsymbol{f}$ and $\boldsymbol{G}$ can be obtained by replacing occurences of $\|\cdot\|_2$ with its interval extension given in Example~\ref{ex:norm-2-extension}. In the rest of the paper, when the input of $\boldsymbol{f}$ and $\boldsymbol{G}$ are intervals, we use such interval extensions.


Finally, we develop Algorithm~\ref{algo:overapprox-datapoints} that utilizes the trajectory $\mathscr{T}_N$ to compute the set $\mathscr{E}_N$ required in Lemma~\ref{lem:overapprox-f-G} to over-approximate $f$ and $G$. Then, we construct a differential inclusion containing the unknown vector field.

\begin{theorem}[\textsc{Differential inclusion}]\label{thm:diff-inclusion}
    \myet{Given a trajectory $\mathscr{T}_N$, the bounds $L_f$ and $L_G$, and the noise bound $\bar{\eta}$}, the dynamics~\eqref{eq:control-linearization} are contained in the differential inclusion
    \begin{align}
        \dot{x} \in \boldsymbol{f}(x) + \boldsymbol{G}(x) u, \label{eq:diff-inclusion}
    \end{align}
    where the functions $\boldsymbol{f} : \mathcal{X} \to \mathbb{IR}^{n}$ and $\boldsymbol{G} : \mathcal{X} \to \mathbb{IR}^{n \times m}$ are obtained by~\eqref{eq:overapprox-f-G} with $\mathscr{E}_N$ being the output of Algorithm~\ref{algo:overapprox-datapoints}.
\end{theorem}

A proof of Theorem~\ref{thm:diff-inclusion} is provided in our previous work~\cite{djeumou2020onthefly}, where we show that the output $\mathscr{E}_N$ of Algorithm~\ref{algo:overapprox-datapoints} is such that for all \myet{$(\Tilde{x}_i, C_{\mathcal{F}_i}, C_{\mathcal{G}_i}) \in \mathscr{E}_N$, $f(\Tilde{x}_i) \in C_{\mathcal{F}_i}$ and $G(\Tilde{x}_i) \in C_{\mathcal{G}_i}$}. Therefore, Lemma~\ref{lem:overapprox-f-G} enables to conclude.

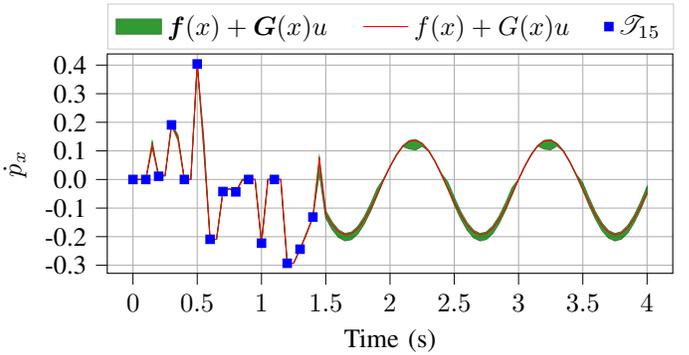
\begin{figure}[!hbt]
    \vspace*{-1.5mm}
    \centering
    \hspace*{-0.4cm}
\begin{tikzpicture}

\definecolor{color0}{rgb}{1,0,0}

\begin{axis}[
width=9.1cm,
height=4.5cm,
legend cell align={left},
legend columns=4,
legend style={fill opacity=1.0, draw opacity=1, text opacity=1, at={(-0.0,1.02)}, anchor=south west, draw=white!80!black, /tikz/every even column/.append style={column sep=0.4cm}},
tick align=outside,
tick pos=left,
x grid style={white!69.0196078431373!black},
xlabel={Time (s)},
xmajorgrids,
xmin=-0.2, xmax=4.2,
xtick style={color=black},
y grid style={white!69.0196078431373!black},
ylabel={\(\displaystyle \dot{p}_x\)},
ymajorgrids,
ymin=-0.328158420875654, ymax=0.438957878954998,
ytick style={color=black},
ytick={-0.4,-0.3,-0.2,-0.1,0,0.1,0.2,0.3,0.4,0.5},
yticklabels={-0.4,-0.3,-0.2,-0.1,0.0,0.1,0.2,0.3,0.4,0.5}
]
\path [draw=green!39.2156862745098!black, fill=green!50.1960784313725!black, opacity=0.8]
(axis cs:0,1.21702228017461e-08)
--(axis cs:0,-1.21702229527027e-08)
--(axis cs:0.05,-1.21702229527027e-08)
--(axis cs:0.1,-8.62955087834588e-09)
--(axis cs:0.15,0.111438585617043)
--(axis cs:0.2,0.0102069063207028)
--(axis cs:0.25,0.0120295121519494)
--(axis cs:0.3,0.18709215279122)
--(axis cs:0.35,0.137236237772356)
--(axis cs:0.4,0)
--(axis cs:0.45,-1.67492502090028e-10)
--(axis cs:0.5,0.404088953765431)
--(axis cs:0.55,0.102691563608646)
--(axis cs:0.6,-0.209187795414084)
--(axis cs:0.65,-0.209187808953117)
--(axis cs:0.7,-0.0421936372534918)
--(axis cs:0.75,-0.0328771349306941)
--(axis cs:0.8,-0.0439868950840471)
--(axis cs:0.85,-0.00359106946594298)
--(axis cs:0.9,0)
--(axis cs:0.95,-0.00036360608281823)
--(axis cs:1,-0.223393414691397)
--(axis cs:1.05,-0.00176455474717177)
--(axis cs:1.1,0)
--(axis cs:1.15,-7.70344255347512e-05)
--(axis cs:1.2,-0.293289494721315)
--(axis cs:1.25,-0.293289498156079)
--(axis cs:1.3,-0.244092759593384)
--(axis cs:1.35,-0.191960237328251)
--(axis cs:1.4,-0.131688091272526)
--(axis cs:1.45,0.0213076580687071)
--(axis cs:1.5,-0.134837385291001)
--(axis cs:1.55,-0.175294005770921)
--(axis cs:1.6,-0.202637102487578)
--(axis cs:1.65,-0.214528257295698)
--(axis cs:1.7,-0.210232320843511)
--(axis cs:1.75,-0.190072260720353)
--(axis cs:1.8,-0.15533921717713)
--(axis cs:1.85,-0.10985004402808)
--(axis cs:1.9,-0.0573209445980477)
--(axis cs:1.95,-0.00239548353804621)
--(axis cs:2,0.0434390576037566)
--(axis cs:2.05,0.0840125196707366)
--(axis cs:2.1,0.115561440018338)
--(axis cs:2.15,0.106779215742591)
--(axis cs:2.2,0.103331835311883)
--(axis cs:2.25,0.116346571587446)
--(axis cs:2.3,0.0994775126961574)
--(axis cs:2.35,0.0615271799568313)
--(axis cs:2.4,0.0150920177417084)
--(axis cs:2.45,-0.0344286752309975)
--(axis cs:2.5,-0.0888060464347705)
--(axis cs:2.55,-0.137389882744605)
--(axis cs:2.6,-0.177207082593946)
--(axis cs:2.65,-0.203719997689711)
--(axis cs:2.7,-0.214718709101182)
--(axis cs:2.75,-0.209513316953042)
--(axis cs:2.8,-0.188479326064832)
--(axis cs:2.85,-0.153027836375142)
--(axis cs:2.9,-0.107077556672302)
--(axis cs:2.95,-0.0542378113558116)
--(axis cs:3,-0.00067201881300404)
--(axis cs:3.05,0.0462169156150209)
--(axis cs:3.1,0.0862122360645805)
--(axis cs:3.15,0.117036571281015)
--(axis cs:3.2,0.106149085024091)
--(axis cs:3.25,0.103630059069684)
--(axis cs:3.3,0.117546510456793)
--(axis cs:3.35,0.0974953534417788)
--(axis cs:3.4,0.0589163439575944)
--(axis cs:3.45,0.0121519769475706)
--(axis cs:3.5,-0.0374823362716509)
--(axis cs:3.55,-0.0917005547005161)
--(axis cs:3.6,-0.139912126737054)
--(axis cs:3.65,-0.179078647190371)
--(axis cs:3.7,-0.204754715176099)
--(axis cs:3.75,-0.214857873620087)
--(axis cs:3.8,-0.208744497438418)
--(axis cs:3.85,-0.18684169210509)
--(axis cs:3.9,-0.150681778986213)
--(axis cs:3.95,-0.10428390174331)
--(axis cs:4,-0.0511481690857607)
--(axis cs:4,-0.0258357545967129)
--(axis cs:4,-0.0258357545967129)
--(axis cs:3.95,-0.0786618942520206)
--(axis cs:3.9,-0.125231352198916)
--(axis cs:3.85,-0.160739735351686)
--(axis cs:3.8,-0.182804275525524)
--(axis cs:3.75,-0.189029823747224)
--(axis cs:3.7,-0.178741286249218)
--(axis cs:3.65,-0.153116565120116)
--(axis cs:3.6,-0.11465574773803)
--(axis cs:3.55,-0.0658476601075241)
--(axis cs:3.5,-0.012320372117535)
--(axis cs:3.45,0.0171194598687223)
--(axis cs:3.4,0.0613764768811685)
--(axis cs:3.35,0.100617336052529)
--(axis cs:3.3,0.12531554498627)
--(axis cs:3.25,0.139231996373379)
--(axis cs:3.2,0.136712970418972)
--(axis cs:3.15,0.11968112684188)
--(axis cs:3.1,0.0891271889986845)
--(axis cs:3.05,0.0489560904418608)
--(axis cs:3,0.00067203470212921)
--(axis cs:2.95,-0.0288697276366728)
--(axis cs:2.9,-0.0815068089044805)
--(axis cs:2.85,-0.12753513828354)
--(axis cs:2.8,-0.162347862393236)
--(axis cs:2.75,-0.183587201820233)
--(axis cs:2.7,-0.188888105750908)
--(axis cs:2.65,-0.177687583120878)
--(axis cs:2.6,-0.151278722408312)
--(axis cs:2.55,-0.112178949583283)
--(axis cs:2.5,-0.0629000415983703)
--(axis cs:2.45,-0.0092106805990715)
--(axis cs:2.4,0.0201134463655048)
--(axis cs:2.35,0.0639402707903291)
--(axis cs:2.3,0.102588144060611)
--(axis cs:2.25,0.126515483855617)
--(axis cs:2.2,0.13953022013118)
--(axis cs:2.15,0.136082839700472)
--(axis cs:2.1,0.118178928950529)
--(axis cs:2.05,0.0868871108361463)
--(axis cs:2,0.046228283926295)
--(axis cs:1.95,0.00239549942717138)
--(axis cs:1.9,-0.0318973089285982)
--(axis cs:1.85,-0.084330167587887)
--(axis cs:1.8,-0.129804872584591)
--(axis cs:1.75,-0.163912095523432)
--(axis cs:1.7,-0.184319398442637)
--(axis cs:1.65,-0.188694159416883)
--(axis cs:1.6,-0.176584818282008)
--(axis cs:1.55,-0.149400115437953)
--(axis cs:1.5,-0.109672443074789)
--(axis cs:1.45,0.0769917274128368)
--(axis cs:1.4,-0.130902517030618)
--(axis cs:1.35,-0.185523165815783)
--(axis cs:1.3,-0.243502738398967)
--(axis cs:1.25,-0.292639988158051)
--(axis cs:1.2,-0.292639991592815)
--(axis cs:1.15,7.70453180314179e-05)
--(axis cs:1.1,0)
--(axis cs:1.05,0.00176455647497999)
--(axis cs:1,-0.222058019086644)
--(axis cs:0.95,0.000363618226714202)
--(axis cs:0.9,0)
--(axis cs:0.85,0.00117522786092749)
--(axis cs:0.8,-0.0425424288352731)
--(axis cs:0.75,-0.0313931872045612)
--(axis cs:0.7,-0.0421936103471051)
--(axis cs:0.65,-0.209187754576542)
--(axis cs:0.6,-0.209187768115574)
--(axis cs:0.55,0.14785660295264)
--(axis cs:0.5,0.404088956235423)
--(axis cs:0.45,1.67492502090028e-10)
--(axis cs:0.4,0)
--(axis cs:0.35,0.155476605034121)
--(axis cs:0.3,0.194615129483986)
--(axis cs:0.25,0.0158125381876669)
--(axis cs:0.2,0.012231731426306)
--(axis cs:0.15,0.132271196038889)
--(axis cs:0.1,1.54916072596034e-08)
--(axis cs:0.05,1.21702228017461e-08)
--(axis cs:0,1.21702228017461e-08)
--cycle;
\addlegendimage{area legend, draw=green!39.2156862745098!black, fill=green!50.1960784313725!black, opacity=0.8}
\addlegendentry{$\boldsymbol{f}(x) + \boldsymbol{G}(x) u$}

\addplot [color0]
table {%
0 -1.33484622222029e-16
0.05 -1.33484622222029e-16
0.1 3.43102819062874e-09
0.15 0.119543134412317
0.2 0.0112193188735044
0.25 0.0137057935754986
0.3 0.190853641137603
0.35 0.147986905693413
0.4 0
0.45 0
0.5 0.404088955000427
0.55 0.133928395834414
0.6 -0.209187781764829
0.65 -0.209187793992728
0.7 -0.0421936238002985
0.75 -0.0325629482989651
0.8 -0.0432646619596601
0.85 -0.00323545432550594
0.9 0
0.95 0
1 -0.222725712727838
1.05 -0.00159621956901854
1.1 0
1.15 0
1.2 -0.292964738995882
1.25 -0.292964742111426
1.3 -0.243797748996176
1.35 -0.191212738674601
1.4 -0.131295304151572
1.45 0.0699365901055328
1.5 -0.122249552403653
1.55 -0.158334382146164
1.6 -0.18275595522157
1.65 -0.193470452403528
1.7 -0.18960219882203
1.75 -0.171465758798659
1.8 -0.140559240026886
1.85 -0.0995140374974853
1.9 -0.0519456518104482
1.95 -0.00215809084540991
2 0.0452888679952687
2.05 0.0860720186473361
2.1 0.116548106681572
2.15 0.134067735033843
2.2 0.137144793470629
2.25 0.125521434316735
2.3 0.100176726694131
2.35 0.0632895544793893
2.4 0.0181192152152391
2.45 -0.0312462740371442
2.5 -0.0802860993472515
2.55 -0.124531526426962
2.6 -0.16003589786429
2.65 -0.183732614793123
2.7 -0.193641874613213
2.75 -0.188954486176323
2.8 -0.170051759820502
2.85 -0.138497284492794
2.9 -0.0969827777916704
2.95 -0.0491716074777475
3 0.00060542148349361
3.05 0.0477887305414651
3.1 0.0880824411532309
3.15 0.117892108086405
3.2 0.134630273766364
3.25 0.13687865713229
3.3 0.124448885108221
3.35 0.098388787452636
3.4 0.06094199255238
3.45 0.0154223102450682
3.5 -0.0340463037067454
3.55 -0.0829327529651172
3.6 -0.126785806562418
3.65 -0.161700032082953
3.7 -0.184665653634183
3.75 -0.193767130330292
3.8 -0.188261806312089
3.85 -0.168597717585803
3.9 -0.136403780313823
3.95 -0.0944315576398158
4 -0.0463913261092619
};
\addlegendentry{$f(x) + G(x) u$}
\addplot [semithick, blue, mark=square*, mark size=1.5, mark options={solid}, only marks]
table {%
0 -1.33484622222029e-16
0.1 3.43102819062874e-09
0.2 0.0112193188735044
0.3 0.190853641137603
0.4 0
0.5 0.404088955000427
0.6 -0.209187781764829
0.7 -0.0421936238002985
0.8 -0.0432646619596601
0.9 0
1 -0.222725712727838
1.1 0
1.2 -0.292964738995882
1.3 -0.243797748996176
1.4 -0.131295304151572
};
\addlegendentry{$\mathscr{T}_{15}$}
\end{axis}

\end{tikzpicture}
    \vspace*{-5mm}
    \caption{Time evolution of $\dot{p}_x$ and its over-approximation $(\boldsymbol{f}(x) + \boldsymbol{G}(x) u)_1$ for the  unicycle system of Example~\ref{ex:unicycle-system} \myet{in a noiseless setting.} We generate the trajectory corresponding to $\dot{x}(t)$ using a randomly generated piecewise-constant input $u(t) \in \mathcal{U}$ for $t \leq t_{15} = 1.5s$, and $u(t) = [1 , \cos(6(t-t_{15}))]$ for $t \in [t_{15}, 4]$.}
    \vspace*{-3mm}
    \label{fig:diff-inclusion-unicycle}
\end{figure}

Figure~\ref{fig:diff-inclusion-unicycle} shows that the differential inclusion holds on the unicycle system using a randomly generated trajectory $\mathscr{T}_{15}$.

\begin{remark}[\textsc{Persistent Excitation}]
    The quality of the differential inclusion of Theorem~\ref{thm:diff-inclusion} depends on how much information on $f$ and $G$ can be obtained from $\mathscr{T}_N$. Thus, if the trajectory is not \emph{diverse}, it is likely impossible to obtain tight differential inclusions. For example, consider one of the corner cases where $u_i =0$ for every data point in $\mathscr{T}_N$. In this case, it is impossible to retrieve any information on $G$. When the goal is to control the system, control values can be synthesized to diversify the trajectory and obtain tight differential inclusions that help future control decisions. For example, excitation-based control (i.e., control inputs that are either zero or nonzero only on a single axis) can be performed to learn the dynamics. 
\end{remark}

\subsection{Bound on the Time Step Size for the Existence of an a Priori Rough Enclosure}
In this section, we consider the problem of finding an a priori rough enclosure $\mathcal{S}_i$ solution of the fixed-point equation
\begin{align}
    \mathcal{R}_{i}^+ \ + \  [0, \Delta t]\ \mathscr{R}(h, \mathcal{S}_{i} \times \boldsymbol{v}([t_i,t_i+\Delta t])) \ \subseteq \ \mathcal{S}_{i}, \label{eq:apriori-enclosure-control-affine}
\end{align}
where $\mathscr{R}(\cdot, \cdot)$ defined in~\eqref{eq:range-def-over-approximation} represents the range of a function over an interval domain, $\mathcal{R}_i^+ \in \mathbb{IR}^n$ is the set of states at time $t_i \geq 0$, $\Delta t > 0$ is the time step size, $\boldsymbol{v}([t_i,t_i+\Delta t]) \in \mathbb{IR}^m$ over-approximates the range of all $u \in \mathbb{V}$ on the interval $[t_i, t_i +\Delta t]$ for a given set $\mathbb{V} \subseteq \mathbb{U}$, and the unknown function $h$ is given by $h(x,u) = f(x) + G(x) u$ for $x \in \mathcal{X}, u \in \mathcal{U}$. Specifically, we give a bound on the time step size $\Delta t$ such that $\mathcal{S}_i$ exists and provide an analytic expression for $\mathcal{S}_i$. Recall that $\mathcal{S}_i$ and $\Delta t$ enable to compute an over-approximation of the reachable set when using interval Taylor-based methods.

\begin{theorem}[\textsc{Explicit a priori rough enclosure}] \label{thm:step-size-fixpoint}
    Assume that the time step size $\Delta t$ satisfies
    \begin{align} \label{eq:step-size-bound}
        (\sqrt{n} \beta_i)\Delta t < 1,
    \end{align}
    \myet{where $\beta_i = \sqrt{ \sum_{k=1}^n \Big( L_{f_k} + \sum_{l=1}^m L_{G_{k,l}} |(\boldsymbol{v}([t_i,t_i+\Delta t]))_l| \Big)^2 }$. Then, the set $\mathcal{S}_i$ given by}
    \begin{align}
        \mathcal{S}_i = \mathcal{R}_i^+ +  \frac{\Delta t \|\boldsymbol{f}(\mathcal{R}_i^+) + \boldsymbol{G}(\mathcal{R}_i^+) \boldsymbol{v}([t_i,t_i+\Delta t])\|_{\infty}}{1- \sqrt{n} \Delta t \beta_i} [-1,1]^n \label{eq:explict-fixpoint}
    \end{align}
    is an a priori rough enclosure solution of the fixed-point equation~\eqref{eq:apriori-enclosure-control-affine}. The functions $\boldsymbol{f}$ and $\boldsymbol{G}$ in~\eqref{eq:explict-fixpoint} can be any known interval extensions of the unknown functions $f$ and $G$.
\end{theorem}

A proof of Theorem~\ref{thm:step-size-fixpoint} is \myet{based on several lemmas and is provided in Appendix~\ref{appendix-A}. Intuitively, we use the bounds on the Lipschitz constants to provide explicit bounds on the rate of changes of solutions of the differential equation~\eqref{eq:control-linearization}, which provide intuition on how to pick a set $\mathcal{S}_i$. However, the condition involves $\beta_i$ which is also dependent on $\Delta t$, as $\boldsymbol{v}([t_i, t_i +\Delta t])$ depends on $\Delta t$. The following corollary gives an explicit bound on $\Delta t$ based on the constraint set $\mathcal{U}$.}


\begin{corollary}\label{corr:step_size_bounds}
    Under the notation of Theorem~\ref{thm:step-size-fixpoint}, assume that 
    \begin{align}
        (\sqrt{n} \beta_{\infty})\Delta t < 1, \label{eq:inf-bounds-step-size}
    \end{align}
    where $\beta_{\infty} = \sqrt{ \sum_{k=1}^n (L_{f_k} + \sum_{l=1}^m L_{G_{k,l}} |{(\mathcal{U})}_l|)^2 }$. Then, $\mathcal{S}_i$ given by~\eqref{eq:explict-fixpoint} is a solution of~\eqref{eq:apriori-enclosure-control-affine}.
\end{corollary}

\begin{proof}
    This is immediate by observing that $\frac{1}{\beta_{\infty}} \leq \frac{1}{\beta_i}$ since $\boldsymbol{v}([t_i,t_i+\Delta t]) \subseteq \mathcal{U}$. Hence, Theorem~\ref{thm:step-size-fixpoint} applies.
\end{proof}

\subsection{Interval Taylor-Based Method for Differential Inclusions}

We compute an over-approximation of the reachable set of the dynamics described by the differential inclusion~\eqref{eq:diff-inclusion}. The obtained over-approximation is also an over-approximation of the reachable set of the unknown system. Theorem~\ref{thm:over-approximation-state} provides a closed-form expression for such over-approximating sets.
\begin{theorem}[\textsc{Reachable set over-approximation}] \label{thm:over-approximation-state}
    Given a trajectory $\mathscr{T}_N$, a set $\mathbb{V} \subseteq \mathbb{U}$ of \emph{piecewise-}$\mathscr{C}^{D_u}$ control signals for $D_u \geq  1$, the bounds $L_f$ and $L_G$, the set $\mathcal{R}^+_i \in \mathbb{IR}^n$ of states at time $t_i$, assume that $\Delta t > 0$ satisfies~\eqref{eq:step-size-bound}. Then, an over-approximation $\mathcal{R}^+_{i+1}$ of the reachable set at $t_i+ \Delta t$ of dynamics described by the differential inclusion $\dot{x}\in \boldsymbol{f}(x) + \boldsymbol{G}(x) u$ with $u\in\mathbb{V}$ is given by
    \begin{align} 
        \mathcal{R}^+_{i+1} = \begin{aligned}[t] 
                                    \mathcal{R}^+_i &+  \Big( \boldsymbol{f}(\mathcal{R}^+_i) + \boldsymbol{G} (\mathcal{R}^+_i) \boldsymbol{v}(t_i) \Big) \Delta t  \\
                                    &+  \Big( \mathcal{J}_f + \mathcal{J}_G \mathcal{V}_i \Big) \Big( \boldsymbol{f}(\mathcal{S}_i) + \boldsymbol{G}(\mathcal{S}_i) \mathcal{V}_i \Big) \frac{\Delta t^2}{2} \\
                                    &+  \boldsymbol{G}(\mathcal{S}_i) \mathcal{V}^{(1)}_i \frac{\Delta t^2}{2},
                                \end{aligned} \label{eq:over-approx-next-state}
    \end{align}
    where $\boldsymbol{v}(t_i)$, $\mathcal{V}_i = \boldsymbol{v}([t_i, t_i+\Delta t])$, and $\mathcal{V}^{(1)}_i = \boldsymbol{v}^{(1)}([t_i, t_i+\Delta t])$ denote the intervals that contain the range of all the control signals in $\mathbb{V}$ and their first derivative on $[t_i, t_i+ \Delta t]$ and at $t_i$. The interval matrices $\mathcal{J}_f \in \mathbb{IR}^{n \times n}$ and $\mathcal{J}_G \in \mathbb{IR}^{n \times m \times n}$ are interval extensions of the Jacobian of $f$ and $G$, given by
    \begin{align}
        \mathcal{J}_{f}&= \left[\begin{array}{ccc}
                L_{f_{1}}[-1, 1] & \cdots &L_{f_{1}}[-1, 1]\\
                \vdots &\ddots &\vdots\\
                L_{f_{n}}[-1, 1] & \cdots &L_{f_{n}}[-1, 1]\\
        \end{array}\right],\label{eq:over-approx-jac-f} \\
        \mathcal{J}_G &= \left[\begin{array}{ccc}
                L_{G_{1,1}}{[-1,1]}^n & \cdots &L_{G_{1,m}}{[-1,1]}^n\\
                \vdots &\ddots &\vdots\\
                L_{G_{n,1}}{[-1,1]}^n & \cdots &L_{G_{n,m}}{[-1,1]}^n\\
        \end{array}\right], \label{eq:over-approx-jac-gkl}
    \end{align}
    and the set $\mathcal{S}_i$ can be either obtained by~\eqref{eq:explict-fixpoint} of Theorem~\ref{thm:step-size-fixpoint} or by solving the fixed-point equation
    \begin{equation} \label{eq:fix-point-rough}
        \mathcal{R}^+_i \ +  \ [0, \Delta t] \ \Big( \boldsymbol{f}(\mathcal{S}_i) + \boldsymbol{G}(\mathcal{S}_i) \mathcal{V}_i \Big) \ \subseteq \ \mathcal{S}_i.
    \end{equation}
    For $k,p\in \mathbb{N}_{[1,n]}$, we define the $(k,p)$ component of $\mathcal{J}_G \mathcal{V}_i$ as
\begin{align}
    {(\mathcal{J}_G \mathcal{V}_i)}_{k,p} &= \sum_{l=1}^m
                {(\mathcal{J}_G)}_{k,l,p}
    {(\mathcal{V}_i)}_l. \label{eq:over-approx-jac-Gu}
\end{align}
\end{theorem}

A proof of Theorem~\ref{thm:over-approximation-state} is provided in our previous work~\cite{djeumou2020onthefly}, where we use a Taylor expansion~\eqref{eq:taylor-expansion} of order $D=2$ with an over-approximation of the unknown Jacobian of $f$ and $G$ obtained from the Lipschitz bounds. Besides, Theorem~\ref{thm:step-size-fixpoint} ensures the existence of an a priori rough enclosure $\mathcal{S}_i$.

\begin{remark}
    Theorem~\ref{thm:over-approximation-state} implicitly relies on the knowledge of the intervals $\boldsymbol{v}(t_i)$, $\boldsymbol{v}([t_i, t_i+\Delta t])$, and $\boldsymbol{v}^{(1)}([t_i, t_i+\Delta t])$ given solely by the definition of the set $\mathbb{V}$. For such intervals to exist, all the controls $u \in \mathbb{V}$ must be at least $\mathscr{C}^1$ on the subdomain $[t_i, t_i +\Delta t]$. Hence, the assumption that the controls in $\mathbb{V}$ are \emph{piecewise-}$\mathscr{C}^{D_u}$ with $D_u \geq 1$. However, when $\mathbb{V}$ is such that $D_u = 0$, we propose Corollary~\ref{corr:non-c1-control} to over-approximate the reachable set of the system. We will focus on Theorem~\ref{thm:over-approximation-state} in the remainder of the paper since, in practise, the control signals of interest are piecewise-constants.\vspace*{-1mm}
\end{remark}

\algdef{SE}[DOWHILE]{Do}{doWhile}{\algorithmicdo}[1]{\algorithmicwhile\ #1}%
\begin{algorithm}[!t]
    \caption{\reachalgo{}: Over-approximation of the reachable set of unknown smooth systems.}\label{algo:DaTaReach}
    \begin{algorithmic}[1]    
        \Require{Single trajectory $\mathscr{T}_N$, upper bounds on the Lipschitz constants, set $\mathbb{V}$ of control signals, time step size $\Delta t$ satisfying~\eqref{eq:step-size-bound} or~\eqref{eq:inf-bounds-step-size}, maximum number of time steps $T > N$, \myet{noise bounds $\eta \geq 0$ and $\bar{\eta} \geq 0$}.\newline
        \emph{Optional}: Any of the side information~\ref{side:bounds-state-vectorfield}--\ref{side:partial-knowledge}.}
    \Ensure{Over-approximations ${\{\mathcal{R}^+_i\}}_{i = N+1}^T$ of the reachable sets at times $t_i=t_N + (i-N)\Delta t$ with $i \in \mathbb{N}_{[N+1,T]}$.} 
        \State \myet{Define $R_{N+1}^+\gets\{[\Tilde{x}_{N+1}-\eta, \Tilde{x}_{N+1} + \eta]\}$}
        \For{$i\in\mathbb{N}_{[N+1,T-1]}$}
                \State Compute $\boldsymbol{v}(t_i)$, $\mathcal{V}_i$, and $\mathcal{V}^{(1)}_i$ in Theorem~\ref{thm:over-approximation-state} from $\mathbb{V}$
                \State Compute $\boldsymbol{f}(\mathcal{R}_i^+)$ and $\boldsymbol{G}(\mathcal{R}_i^+)$ via Theorem~\ref{thm:diff-inclusion}
                \State Get tighter $\boldsymbol{f}(\mathcal{R}_i^+)$ and $\boldsymbol{G}(\mathcal{R}_i^+)$ if any side information
                \State Compute $\mathcal{S}_i$ via~\eqref{eq:explict-fixpoint} or~\eqref{eq:fix-point-rough},  $\boldsymbol{f}(\mathcal{R}_i^+)$, and $\boldsymbol{G}(\mathcal{R}_i^+)$
                \State Compute $\boldsymbol{f}(\mathcal{S}_i)$ and $\boldsymbol{G}(\mathcal{S}_i)$ via Theorem~\ref{thm:diff-inclusion}
                \State Get tighter $\boldsymbol{f}(\mathcal{S}_i)$ and $\boldsymbol{G}(\mathcal{S}_i)$ if any side information
                \State Compute $\mathcal{J}_f$ and $\mathcal{J}_G$ via~\eqref{eq:over-approx-jac-f} and~\eqref{eq:over-approx-jac-gkl}
                \State Get tighter $\mathcal{J}_f$ and $\mathcal{J}_G$ if any side information
                \State Compute $\mathcal{R}^+_{i+1}$ via~\eqref{eq:over-approx-next-state}
        \EndFor
    \State\Return ${\{\mathcal{R}^+_i\}}_{i=N+1}^T$
  \end{algorithmic}
\end{algorithm}

\begin{corollary}\label{corr:non-c1-control}
    Under the notation of Theorem~\ref{thm:over-approximation-state}, assume that $D_u = 0$. Then, an over-approximation $\mathcal{R}_{i+1}^+$ of the reachable set of the dynamics described by the differential inclusion $\dot{x}\in \boldsymbol{f}(x) + \boldsymbol{G}(x) u$ with $u\in\mathbb{V}$ is given by 
    \begin{align*}
        \mathcal{R}_{i+1}^+ = \mathcal{R}_i^+ + \big( \boldsymbol{f}(\mathcal{S}_i) + \boldsymbol{G}(\mathcal{S}_i) \mathcal{V}_i \big) \Delta t.
    \end{align*}
\end{corollary} 

\subsection{\myet{Side Information}}
\myet{As motivated in the introduction, prior knowledge on the underlying unknown dynamics might be available}. However, Theorem~\ref{thm:over-approximation-state}, as it is, does not incorporate any side information. We now show how to incorporate a variety of side information to obtain tighter over-approximations of the reachable sets.\vspace*{2mm}
\begin{figure}[!t]
    \centering
    \hspace*{-0.5cm}
    \input{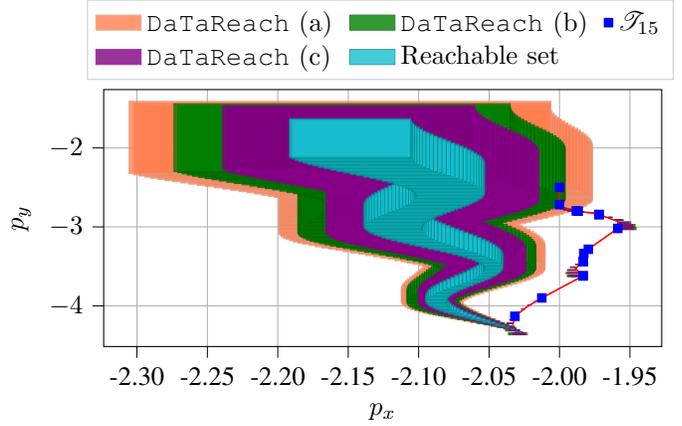}
    \vspace*{-4mm}
    \caption{Over-approximation of the reachable set of the unicycle system in the $x$-$y$ plane. The trajectory $\mathscr{T}_{15}$ comes from a \myet{noise-free} trajectory of the system obtained by a randomly generated piecewise-constant control signal $u(t) \in \mathcal{U}$ for $t \leq t_{15} = 1.5s$. The parameters for \reachalgo{} were given by $\Delta t = 0.02$, $T = 200$, and $\mathbb{V} = \{t \mapsto [1+ a_1, \cos(6(t-t_{15})) + a_2] \: \vert \: a_1 \in [-0.1,0.1], a_2 \in [-0.01,0.01]\}$. The case $\mathrm{(a)}$ ignores the side information $f(x) = f(\theta)$, case $\mathrm{(b)}$ is exactly the setting of Example~\ref{ex:unicycle-system}, and case $\mathrm{(c)}$ assumes the extra knowledge that $f=0$ and $(G)_{3,2} = 1$.}
    \vspace*{-3mm}
    \label{fig:overapprox-unicycle}
\end{figure}



\begin{sideinfo}[\textsc{Vector field bounds}]\label{side:bounds-state-vectorfield}
\myet{We are given  $\mathcal{R}^{f_\mathcal{A}} \in \mathbb{IR}^n$ and $\mathcal{R}^{G_{\mathcal{A}}} \in \mathbb{IR}^{n \times m}$ as supersets of the range of $f$ and $G$, respectively, over a given set $\mathcal{A} \subseteq \mathcal{X}$.}

\myet{Given a set $\mathcal{S} \subseteq \mathcal{A}$, tight extensions of $\boldsymbol{f}$ and $\boldsymbol{G}$ over $\mathcal{S}$ can be obtained by the update $\boldsymbol{f}(\mathcal{S}) \gets \boldsymbol{f}(\mathcal{S}) \cap \mathcal{R}^{f_\mathcal{A}}$ and $\boldsymbol{G}(\mathcal{S}) \gets \boldsymbol{G}(\mathcal{S}) \cap \mathcal{R}^{G_\mathcal{A}}$.}
\end{sideinfo}

\begin{sideinfo}[\textsc{Gradient bounds}]\label{side:gradient-bounds}
    \myet{We are given bounds on the gradient of some components of $f$ and $G$. Such side information may include the monotonicity of $f$ and $G$.}
    
    \myet{These bounds can be used in equations~\eqref{eq:over-approx-jac-f} and~\eqref{eq:over-approx-jac-gkl} to provide tight interval extensions $\mathcal{J}_f$ and $\mathcal{J}_G$. For example, if the function $(f)_k$ is known to be non-decreasing with respect to the variable $x_p$ on a set $\mathcal{A} \subseteq \mathcal{X}$, then we obtain a tighter $\mathcal{R}_{i+1}^+$ by the update $(\mathcal{J}_f)_{k,p} \gets (\mathcal{J}_f)_{k,p} \cap \mathbb{R}_+$ if $\mathcal{S}_i \subseteq \mathcal{A}$.}
    
    \myet{This side information expresses properties such as decoupling between states. For example, if the state ${(x(t))}_{p}$ does not directly affect ${(\dot{x}(t))}_{k}$ for some $p,k \in \mathbb{N}_{[1,n]}$ under any control signal in $ \mathcal{U}$, we can obtain a tighter over-approximation of the reachable set by setting to zero the intervals ${(\mathcal{J}_f)}_{k,p}$ and ${(\mathcal{J}_G)}_{k, l, p}$ for all $l \in \mathbb{N}_{[1,m]}$. For the unicycle system of Example~\ref{ex:unicycle-system}, since $f$ and $G$ depends only on the heading $\theta$, the Jacobian terms $(\mathcal{J}_f)_{1,1}$, $(\mathcal{J}_f)_{1,2}$, $(\mathcal{J}_f)_{2,1}$, $(\mathcal{J}_f)_{2,2}$, $(\mathcal{J}_f)_{3,1}$, $(\mathcal{J}_f)_{3,2}$, ${(\mathcal{J}_G)}_{1,1,1}$, ${(\mathcal{J}_G)}_{1,1,2}$, ${(\mathcal{J}_G)}_{2,1,1}$, ${(\mathcal{J}_G)}_{2,1,2}$, ${(\mathcal{J}_G)}_{3,2,1}$, and ${(\mathcal{J}_G)}_{3,2,2}$ must all be set to \emph{zero}.}
\end{sideinfo}

\begin{sideinfo}[\textsc{Algebraic constraints}]\label{side:algebraic-equation}
    \myet{We are given a differentiable map $g : \mathbb{R}^n \mapsto \mathbb{R}^q$ satisfying the constraint $g(x) \geq 0$ over the state $x$ of the unknown system.}
    
    \myet{Without loss of generality, we consider that $g : \mathbb{R}^n \mapsto \mathbb{R}$. This side information provides bounds on $f$, $G$, $\mathcal{J}_f$, and $\mathcal{J}_G$ \emph{locally}. Specifically, if we derivate the function $t \mapsto g(x(t))$, then we obtain a new constraint $w : \mathbb{R}^n \times \mathbb{R}^{n \times m} \times \mathbb{R}^m \times \mathbb{R}^n \mapsto \mathbb{R}$ such that $w(f(x), G(x), u, x) \geq 0$. If we compute the partial derivatives of $w$ with respect to all the states $x_i$, then we obtain a new constraint $z : \mathbb{R}^{n \times n} \times \mathbb{R}^{n \times m \times n} \times \mathbb{R}^m \times \mathbb{R}^n \mapsto \mathbb{R}^n$ such that $z(\frac{\partial f}{\partial x}(x), \frac{\partial G}{\partial x}(x), u, x) \geq 0$. The obtained constraints $w$ and $z$ can be incorporated in the computation of $\mathcal{R}_{i+1}^+$ through the framework of contractors~\cite{CHABERT20091079,benhamou-contractor1999}. Specifically, given a set $\mathcal{A}$ of states,  a set $\mathcal{V}$ of control values, and the over-approximating sets $\boldsymbol{f}(\mathcal{A})$, $\boldsymbol{G}(\mathcal{A})$, $\mathcal{J}_f$, and $\mathcal{J}_G$, the methods based on contractor programming~\cite{CHABERT20091079} compute the intervals $C_{\boldsymbol{f}(\mathcal{A})} \subseteq \boldsymbol{f}(\mathcal{A})$, $C_{\boldsymbol{G}(\mathcal{A})} \subseteq \boldsymbol{G}(\mathcal{A})$, $C_{\mathcal{J}_f} \subseteq \mathcal{J}_f$, and $C_{\mathcal{J}_G} \subseteq \mathcal{J}_G$. These intervals are such that for all $f_x \in \boldsymbol{f}(\mathcal{A}) \setminus C_{\boldsymbol{f}(A)}$, $G_x \in \boldsymbol{f}(\mathcal{A}) \setminus C_{\boldsymbol{G}(A)}$, $p_f \in \mathcal{J}_f \setminus C_{\mathcal{J}_f}$, and  $p_G \in \mathcal{J}_G \setminus C_{\mathcal{J}_G}$, there does not exist $x \in \mathcal{A}$ and $u \in \mathcal{V}$ such that $w(f_x,G_x,u,x) \geq 0$ and $z(p_f, p_G, u, x) \geq 0$. In other words, it contracts $\boldsymbol{f}(\mathcal{A})$, $\boldsymbol{G}(\mathcal{A})$, $\mathcal{J}_f$, and $\mathcal{J}_G$ by pruning out some values that do not satisfy the constraints imposed by $w$ and $z$. In the computation of $\mathcal{R}_{i+1}^+$, the set $\mathcal{A}$ is either $\mathcal{R}_i^+$ or $\mathcal{S}_i$, and the set $\mathcal{V}$ is $\mathcal{V}_i$.}
\end{sideinfo}

\begin{sideinfo}[\textsc{Partial dynamics knowledge}]\label{side:partial-knowledge}
    \myet{We are given terms of some components of the vector field of~\eqref{eq:control-linearization}. That is, we consider that the dynamics are in the form $$\dot{x} = f_{\mathrm{kn}}(x) + f_{\mathrm{ukn}}(x) + (G_{\mathrm{kn}}(x) + G_{\mathrm{ukn}}(x)) u,$$ where  the functions $f_{\mathrm{kn}}$ and $G_{\mathrm{kn}}$ are known while $f_{\mathrm{ukn}}$ and $G_{\mathrm{ukn}}$ are unknown. Such side information is usually obtained from the application of elementary laws of physics.}
    
    \myet{The new functions $\boldsymbol{f}$ and $\boldsymbol{G}$ are given by $\boldsymbol{f} = \boldsymbol{f}_{\mathrm{kn}} + \boldsymbol{f}_{ukn}$ and $\boldsymbol{G} = \boldsymbol{G}_{\mathrm{kn}} + \boldsymbol{G}_{ukn}$, where the functions $\boldsymbol{f}_{\mathrm{kn}}$ and $\boldsymbol{f}_{ukn}$ are interval extensions of known functions $f_{\mathrm{kn}}$ and $G_{\mathrm{kn}}$, respectively, and the functions $\boldsymbol{f}_{ukn}$ and $\boldsymbol{G}_{ukn}$ are obtained by Theorem~\ref{thm:diff-inclusion} using $L_{f_{\mathrm{ukn}}}$, $L_{G_{\mathrm{ukn}}}$, and the new trajectory $\mathscr{T}^{'}_{N} = \{(x_i,\dot{x}_i - (f_{\mathrm{kn}}(x_i) +G_{\mathrm{kn}}(x_i) u_i) , u_i) \: \vert \: (x_i, \dot{x}_i, u_i) \in \mathscr{T}_N\}.$
    Furthermore, the new Jacobian terms in the computation of $\mathcal{R}_{i+1}^+$ are given by $\mathcal{J}_f = \boldsymbol{\frac{\partial f_{\mathrm{kn}}}{\partial x}}(\mathcal{S}_i) + \mathcal{J}_{f_{\textrm{ukn}}}$ and $(\mathcal{J}_{G})_{k,l,p} = \boldsymbol{\frac{\partial (G_{\mathrm{kn}})_{k,l}}{\partial x_p}}(\mathcal{S}_i) + (\mathcal{J}_{G_{\textrm{ukn}}})_{k,l,p}$, respectively.}
\end{sideinfo}

\reachalgo{} (Algorithm~\ref{algo:DaTaReach}) addresses Problem~\ref{prob:reachability} by combining Theorem~\ref{thm:over-approximation-state} and the available side information.
We demonstrate the value of additional side information on the unicycle system. Figure~\ref{fig:overapprox-unicycle} shows that, as expected, the over-approximation becomes tighter with richer side information.


\section{Control Synthesis for Unknown Dynamics}

We develop \controlalgo{} to address Problem~\ref{prob:control}. It computes approximate solutions to the one-step optimal control problem~\eqref{eq:one-step-optimal-control} using the over-approximation of the reachable set. Specifically, it replaces the \emph{unknown} next state $x_{i+1}$ in problem~\eqref{eq:one-step-optimal-control} with the corresponding over-approximation of the reachable set $\mathcal{R}^+_{i+1}$, which may be computed using Theorem~\ref{thm:over-approximation-state}. Unfortunately, $\mathcal{R}^+_{i+1}$ is a nonconvex function of the control signal $u$, the decision variable. Therefore, it constructs an over-approximation of $\mathcal{R}^+_{i+1}$, in Theorem~\ref{thm:linear-expr-O}, that is convex in the given constant control signal $u$ applied in the time interval $[t_i,t_{i+1}]$. The new over-approximation provides tractable approximations of the optimization problem \eqref{eq:one-step-optimal-control}.

\subsection{Approximate One-Step Optimal Control}
First, Theorem~\ref{thm:linear-expr-O} provides a control-affine linearization of the over-approximation $\mathcal{R}_{i+1}^+$ given by Theorem~\ref{thm:over-approximation-state}.

\begin{theorem} \label{thm:linear-expr-O}
    Under the notation of Theorem~\ref{thm:over-approximation-state} and given a constant control signal $u : t \mapsto u_i$ applied between $t_i$ and $t_{i+1}=t_i + \Delta t$ with $u_i \in \mathcal{U}$, the reachable set $\mathcal{R}_{i+1}$ satisfies
    \begin{equation}
        \mathcal{R}_{i+1} \subseteq \big(\mathcal{B}_i + \mathcal{A}_i^+ u_i\big) \cap \big(\mathcal{B}_i + \mathcal{A}_i^- u_i\big), \label{eq:linear-over-approx}
    \end{equation}
    where the intervals $\mathcal{A}^-_i$, $\mathcal{A}^+_i$, and $\mathcal{B}_i$ are given by
    \begin{equation} \label{eq:linear-over}
    \begin{aligned}
        \mathcal{A}^-_i &= 
                                \boldsymbol{G}(\mathcal{R}^+_i) \Delta t + \big( \mathcal{J}_f \boldsymbol{G}(\mathcal{S}_i) + \mathcal{J}_G^{\mathrm{T}}( \boldsymbol{f}(\mathcal{S}_i) + \boldsymbol{G}(\mathcal{S}_i)\mathcal{U}) \big) \frac{\Delta t ^2}{2}, \\
        \mathcal{A}^+_i &= 
                                \boldsymbol{G}(\mathcal{R}^+_i) \Delta t + \big( (\mathcal{J}_f + \mathcal{J}_G \mathcal{U}) \boldsymbol{G}(\mathcal{S}_i) + \mathcal{J}_G^{\mathrm{T}} \boldsymbol{f}(\mathcal{S}_i) \big) \frac{\Delta t ^2}{2}, \\
        \mathcal{B}_i &= \mathcal{R}^+_i + \boldsymbol{f} (\mathcal{R}^+_i) \Delta t +   \mathcal{J}_f \boldsymbol{f}(\mathcal{S}_i) \frac{\Delta t^2}{2},
    \end{aligned}
    \end{equation}
    where ${(\mathcal{J}_G^\textrm{T})}_{k,p,l}$ = ${(\mathcal{J}_G)}_{k,l,p}$ for $k,p\in \mathbb{N}_{[1,n]}$ and $l \in \mathbb{N}_{[1,m]}$, and the a priori rough enclosure $\mathcal{S}_i$ is either obtained by~\eqref{eq:explict-fixpoint} with $\boldsymbol{v}([t_i,t_{i+1}])$ replaced by $\mathcal{U}$ or is a solution of
    \begin{equation}\label{eq:apriori-enclosure-control}
        \mathcal{R}^+_i \ + \ [0, \Delta t] \ \Big( \boldsymbol{f}(\mathcal{S}_{i}) + \boldsymbol{G}(\mathcal{S}_{i}) \mathcal{U} \Big) \ \subseteq \ \mathcal{S}_{i}.
    \end{equation}

\end{theorem}

A proof of Theorem~\ref{thm:linear-expr-O} is provided in our previous work~\cite{djeumou2020onthefly}, where we linearize the quadratic term in $u$ of the closed-form expression~\eqref{eq:over-approx-next-state}. Such linearization enables to have a control-affine over-approximation at time $t_i + \Delta t$. \myet{Intuitively, the linearization comes from over-approximating the quadratic term $(\mathcal{J}_G u_i)(\boldsymbol{G}(\mathcal{S}_i) u_i)$ of Theorem~\ref{thm:over-approximation-state} by \emph{linear} terms $(\mathcal{J}_G \mathcal{U})(\boldsymbol{G}(\mathcal{S}_i) u_i)$ and $(\mathcal{J}_G u_i)(\boldsymbol{G}(\mathcal{S}_i) \mathcal{U})$. Thus, we obtain two different over-approximations characterized by $\mathcal{A}^+$ and $\mathcal{A}^-$, which are finally intersected to provide a tighter set.} 

Next, we use the control-affine linearization of Theorem~\ref{thm:linear-expr-O} to propose two convex optimization problems which solutions approximate the one-step optimal control problem~\eqref{eq:one-step-optimal-control}. The first problem, called \emph{optimistic control problem}, is given by
\begin{align}\label{eq:optimistic-control-problem}
    \underset{u_i \in \mathcal{U}}{\mathrm{minimize}} &\quad \underset{ \begin{smallmatrix} x_{i+1} \in \mathcal{X}, \\ x_{i+1} \in (\mathcal{B}_i + \mathcal{A}_i^+ u_i) \cap (\mathcal{B}_i + \mathcal{A}_i^- u_i) \end{smallmatrix}}{\inf} c(\Tilde{x}_i, u_i, x_{i+1}), 
\end{align}
where the goal is to minimize the best possible cost value over all possible state $x_{i+1}$ in the over-approximation~\eqref{eq:linear-over-approx}. The second problem, called \emph{idealistic control problem}, is an idealistic approximation of~\eqref{eq:one-step-optimal-control} given by
\begin{align}\label{eq:idealistic-control-problem}
    \underset{u_i \in \mathcal{U}}{\mathrm{minimize}} &\quad  c(\Tilde{x}_i, u_i, b^{\mathrm{ide}}_i + A^{\mathrm{ide}}_i u_i), 
\end{align}
where the goal is to minimize the cost associated to a specific trajectory $x_{i+1} = b^{\mathrm{ide}}_i + A^{\mathrm{ide}}_i u_i$ in the over-approximation~\eqref{eq:control-linearization}, idealistically considered as the unknown next state evolution. Specifically, we choose $x_{i+1}$ according to the middle point between a point $x_{i+1}^+ \in \mathcal{B}_i + \mathcal{A}^+_i u_i$ and a point $x_{i+1}^- \in \mathcal{B}_i + \mathcal{A}_i^- u_i$. More specifically, given weights $w^+, w^- \in [0,1]$, we pick $b^{\mathrm{ide}}_i$ and $A^{\mathrm{ide}}_i$ as follows
\begin{align}
    b^{\mathrm{ide}}_i &= \frac{1}{2} \Big( w^+ \overline{\mathcal{B}_i} + (1-w^+) \underline{\mathcal{B}_i} + w^- \overline{\mathcal{B}_i} + (1-w^-) \underline{\mathcal{B}_i} \Big), \nonumber \\ 
    A^{\mathrm{ide}}_i &= \frac{1}{2} \Big( w^+ \overline{\mathcal{A}_i^+} + (1-w^+) \underline{\mathcal{A}_i^+} + w^- \overline{\mathcal{A}_i^-} + (1-w^-) \underline{\mathcal{A}_i^-} \Big). \label{eq:coeff-Ai-idealistic-pb}
\end{align}
\begin{remark}
    Another possible approximation of the one-step optimal control problem~\eqref{eq:one-step-optimal-control} is a pessimistic control problem where, in contrast to the optimistic control problem, the goal is to minimize the worst possible one-step cost value over all possible state $x_{i+1}$ in the over-approximation of the reachable set. \myet{Such a pessimistic control problem can be formulated as a semidefinite program using tools from robust optimization.}
\end{remark}    

\algdef{SE}[DOWHILE]{Do}{doWhile}{\algorithmicdo}[1]{\algorithmicwhile\ #1}%
\begin{algorithm}[!t]
    \caption{\controlalgo{}: Approximate one-step optimal control solution at $t_i \geq t_N$}\label{algo:DaTaControl}
    \begin{algorithmic}[1]    
    \Require{ Single trajectory $\mathscr{T}_{i-1}$ of length $i> N$, ime step size $\Delta t$ satisfying~\eqref{eq:inf-bounds-step-size}, Lipschitz bounds, \myet{noise bounds $\eta \geq 0$ and $\bar{\eta} \geq 0$, one-step cost $c$, current state $\Tilde{x}_i = \Tilde{x}(t_{i})$.}\newline
    \emph{Optional}: Any of the side information~\ref{side:bounds-state-vectorfield}--\ref{side:partial-knowledge}.}
    \Ensure{Constant control $\hat{u}_i \in \mathcal{U}$ to apply between $t_i$ and $t_i + \Delta t$ that approximates a solution of~\eqref{eq:one-step-optimal-control}.} 
        \State \myet{Define $\mathcal{R}^+_i \leftarrow \{[\Tilde{x}_i-\eta, \Tilde{x}_i+\eta]\}$} \label{alg:datacontrol-deb}
        \State Compute $\boldsymbol{f}(\mathcal{R}_i^+)$ and $\boldsymbol{G}(\mathcal{R}_i^+)$ via Theorem~\ref{thm:diff-inclusion}
        \State Get tighter $\boldsymbol{f}(\mathcal{R}_i^+)$ and $\boldsymbol{G}(\mathcal{R}_i^+)$ if any side information
        \State Compute $\mathcal{S}_i$ via~\eqref{eq:explict-fixpoint} or~\eqref{eq:apriori-enclosure-control},  $\boldsymbol{f}(\mathcal{R}_i^+)$, and $\boldsymbol{G}(\mathcal{R}_i^+)$
        \State Compute $\boldsymbol{f}(\mathcal{S}_i)$ and $\boldsymbol{G}(\mathcal{S}_i)$ via Theorem~\ref{thm:diff-inclusion}
        \State Get tighter $\boldsymbol{f}(\mathcal{S}_i)$ and $\boldsymbol{G}(\mathcal{S}_i)$ if any side information
        \State Compute $\mathcal{J}_f$ and $\mathcal{J}_G$ via~\eqref{eq:over-approx-jac-f} and~\eqref{eq:over-approx-jac-gkl}
        \State Get tighter $\mathcal{J}_f$ and $\mathcal{J}_G$ if any side information
        \State Compute $\mathcal{B}_{i}$,  $\mathcal{A}^+_{i}$, and $\mathcal{A}^-_{i}$ via~\eqref{eq:linear-over} \label{alg:datacontrol-end}
        \State Compute $\hat{u}_i$ as the solution of either \eqref{eq:optimistic-control-problem} or \eqref{eq:idealistic-control-problem}
        \State\Return $\hat{u}_i$
  \end{algorithmic}
\end{algorithm}

Finally, we provide, in Theorem~\ref{thm:suboptimality-gap}, bounds on the suboptimality of the optimistic and idealistic control with respect to the optimal control when the cost function $c$ is quadratic.

\begin{assumption}[\textsc{Quadratic one-step cost}]\label{ass:quadratic-cost}
    We consider that the one-step cost function $c$ is a convex quadratic function,
    \begin{align}
        c(x,u,y) = \begin{bmatrix} y \\ u\end{bmatrix}^{\mathrm{T}} \begin{bmatrix} Q & S \\ S^{\mathrm{T}}  & R\end{bmatrix} \begin{bmatrix} y \\ u\end{bmatrix} + \begin{bmatrix} q \\ r\end{bmatrix}^{\mathrm{T}} \begin{bmatrix} y \\ u\end{bmatrix},
    \end{align}
    where $q \in \mathbb{R}^n$, $r \in \mathbb{R}^m$, $Q = Q^{\mathrm{T}} \in \mathbb{R}^{n \times n}$, $R = R^{\mathrm{T}} \in \mathbb{R}^{m \times m}$, and $S \in \mathbb{R}^{n \times m}$.
\end{assumption}

\begin{theorem}[\textsc{Suboptimality bound}]\label{thm:suboptimality-gap}
    Let $c_i^{\star}$, $c_i^{\mathrm{opt}}$, and $c_i^{\mathrm{ide}}$ be the optimal cost of the one-step optimal control problem~\eqref{eq:one-step-optimal-control}, the optimistic control problem~\eqref{eq:optimistic-control-problem}, and the idealistic control problem~\eqref{eq:idealistic-control-problem}, respectively, at the sampling time $t_i$. The following inequality holds
    \begin{align}
        |c_i^{\star} - c_i| &\leq \begin{aligned}[t] \max \big( &\| w(\mathcal{B}_i) + w(\mathcal{A}^+_i)|\mathcal{U}| \|_2 K (\mathcal{A}^+_i) , \\ 
        &\| w(\mathcal{B}_i) + w(\mathcal{A}^-_i)|\mathcal{U}| \|_2  K (\mathcal{A}^-_i) \big), \end{aligned} \label{eq:bounds-approximate}
    \end{align}
    where $c_i$ is either $c_i^{\mathrm{opt}}$ or $c_i^{\mathrm{ide}}$, $w(\mathcal{A}) = \overline{\mathcal{A}} - \underline{\mathcal{A}}$ is the width of an interval $\mathcal{A}$, and $K(\mathcal{A})$, for any $\mathcal{A} \in \mathbb{IR}^{n \times m}$, is given by
    \begin{align}
        K(\mathcal{A}) &= \begin{aligned}[t] \min \big( &\| 2 |S \mathcal{U}| + q +  2 | Q (\mathcal{B}_i + \mathcal{A} \mathcal{U} )| \|_2, \\ &\| 2|S \mathcal{U}| + q +  2| Q \mathcal{X}|\|_2 \big). \end{aligned} \nonumber
    \end{align}
\end{theorem}

We provide a proof for Theorem~\ref{thm:suboptimality-gap} in \myet{our previous work~\cite{djeumou2020onthefly}} where we use the local Lipschitz property of the cost function $c$ to characterize its variations over the over-approximation set~\eqref{eq:linear-over-approx}.


\begin{remark}
    The main term of the suboptimality bound in Theorem~\ref{thm:suboptimality-gap} is directly related to the width of the over-approximation of the reachable set at time $t_{i} + \Delta t$. Thus, as the over-approximation of the reachable set becomes tighter, the gap with the unknown optimal cost decreases. Specifically, it is straightforward to see that the more data and richer side information are used in the computation of $\mathcal{B}_i$, $\mathcal{A}^+_i$, and $\mathcal{A}^-_i$, the tighter the widths $w(\mathcal{B}_i)$, $w(\mathcal{A}^+_i)$, and $w(\mathcal{A}^-_i)$ become.
\end{remark}

\subsection{Real-Time Approximate Control Synthesis}
In this section, we investigate efficient algorithms to compute solutions of the optimistic control problem~\eqref{eq:optimistic-control-problem} and the idealistic control problem~\eqref{eq:idealistic-control-problem} when the one-step cost function $c$ satisfies Assumption~\ref{ass:quadratic-cost}. We provide a bound on the number of elementary operations required by \controlalgo{} to synthesize near-optimal control values. Such a bound makes \controlalgo{} a good candidate for real-time control of systems with a priori unknown dynamics.

\begin{lemma}[\textsc{Approximate quadratic problems}] \label{lem:constrained-qp}
    Without loss of generality, consider that either ${(\mathcal{U})}_l \geq 0$ or ${(\mathcal{U})}_l \leq 0$ for a given $l \in \mathbb{N}_{[1,m]}$. The optimistic control problem~\eqref{eq:optimistic-control-problem} can be reformulated as the quadratic programming problem
    \begin{subequations}
        \begin{align}
            \underset{u_i \in \mathcal{U}, x_{i+1}\in \mathcal{X}}{\mathrm{minimize}}&\quad
            c(x_i, u_i, x_{i+1}),\label{eq:optimistic-one-step-cost}\\
            \mathrm{subject\ to}&\quad \underline{\mathcal{B}_i} + A_i^{\mathrm{l}+} u_i \leq x_{i+1} \leq  \overline{\mathcal{B}_i} + A_i^{\mathrm{s}+} u_i\label{eq:approximate_one_step_plus},\\
            &\quad \underline{\mathcal{B}_i} + A_i^{\mathrm{l}-} u_i \leq x_{i+1} \leq  \overline{\mathcal{B}_i} + A_i^{\mathrm{s}-} u_i, \label{eq:approximate_one_step_minus}
        \end{align}\label{eq:optimistic-expr-qp}%
    \end{subequations}
    and the idealistic control problem~\eqref{eq:idealistic-control-problem} as the convex, \emph{box-constrained} quadratic programming problem
    \begin{align}\label{eq:qp-midpoint}
            \underset{u_i \in \mathcal{U}}{\mathrm{minimize}} \quad \frac{1}{2}u_i^{\mathrm{T}} Q_i u_i + q_i^{\mathrm{T}} u_i + p_i,
    \end{align}
    where $Q_i$, $p_i$, and $q_i$ are given by 
    \begin{align}
        Q_i &= 2(A_i^{\mathrm{ide}})^{\mathrm{T}} Q A_i^{\mathrm{ide}} + 4 (A_i^{\mathrm{ide}})^{\mathrm{T}} S + 2R ,\label{eq:Qi_ide}\\
        q_i &= 2\big( S + Q A_i^{\mathrm{ide}} \big)^{\mathrm{T}}b_i^{\mathrm{ide}} + (A_i^{\mathrm{ide}})^{\mathrm{T}} q + r, \label{eq:qi_ide}\\
        p_i &=  (b_i^{\mathrm{ide}})^{\mathrm{T}} Q b_i^{\mathrm{ide}} + q^{\mathrm{T}} b_i^{\mathrm{ide}}, \label{eq:pi-ide}
    \end{align}
    the matrices $A_i^{\mathrm{l}+}$ and $A_i^{\mathrm{s}+}$ are such that, for all $k \in \mathbb{N}_{[1,n]}$ and $l \in \mathbb{N}_{[1,m]}$, if $(\mathcal{U})_l \geq 0$, then $(A_i^{\mathrm{s}+})_{k,l} = \overline{(\mathcal{A}^+_i)_{k,l}}$ and  $(A_i^{\mathrm{l}+})_{k,l} = \underline{(\mathcal{A}^+_i)_{k,l}}$, otherwise $(A_i^{\mathrm{l}+})_{k,l} = \overline{(\mathcal{A}^+_i)_{k,l}}$ and $(A_i^{\mathrm{s}+})_{k,l} = \underline{(\mathcal{A}^+_i)_{k,l}}$, and the matrices $A_i^{\mathrm{l}-}$ and $A_i^{\mathrm{s}-}$ are obtained in a similar manner with $\mathcal{A}^+_i$ replaced by $\mathcal{A}^-_i$.  
\end{lemma}

We provide a proof of Lemma~\ref{lem:bound-general-setting} in Appendix~\ref{appendix-A}, where interval arithmetic and elementary algebra provide~\eqref{eq:optimistic-expr-qp},~\eqref{eq:qp-midpoint}.

\begin{remark}
    Lemma~\ref{lem:constrained-qp} assumes that the components of $\mathcal{U}$ are either nonnengative or nonpositive in order to simplify the notation and provide a simple expression for the optimistic control problem. In the general case, $\mathcal{U}$ can be partitioned into subsets where each component is either nonnegative or nonpositive. Then, we solve the optimistic control problem on each partition and return the control value from the partition with the minimum cost. It is important to note that the idealistic control problem~\eqref{eq:qp-midpoint} does not require any partitioning of $\mathcal{U}$ and can therefore be solved as a one-shot quadratic problem. Besides, if $m_{\mathrm{i}} \leq m$ is the number of components of $\mathcal{U}$ having an indefinite sign, then solving the optimistic control problem~\eqref{eq:optimistic-control-problem} requires to solve $2^{m_{\mathrm{i}}}$ quadratic programming problems similar to~\eqref{eq:optimistic-expr-qp}. As a consequence, the number of problems drastically reduces when the intervals ${(\mathcal{U})}_l$ are known to be nonnegative or nonpositive \emph{a priori}.
\end{remark}

We seek for algorithms to compute solutions of~\eqref{eq:optimistic-expr-qp} and~\eqref{eq:qp-midpoint} while providing bounds on the number of elementary operations to obtain such solutions. The number of elementary operations, referred as flops, is obtained by counting the number of operations $+$, $-$, $*$, $/$, $\max$, $\min$, and $\sqrt{(\cdot)}$ executed by the algorithms. When the cost function $c$ is assumed to be strongly convex, i.e., either $Q$ is positive definite or $R$ is positive definite, the authors of~\cite{patrinos2013accelerated} develop an algorithm~\cite[Algorithm 1]{patrinos2013accelerated} based on applying the fast gradient-projection method on the dual of the optimization problem. The algorithm can be directly used to solve both the optimistic and idealistic control problem. The rate of convergence of such an algorithm is proportional to $1/\sqrt{\epsilon}$, where $\epsilon$ is the desired accuracy for optimality. Furthermore, the authors provide a maximum number of flops, cubic polynomial in both $n$ and $m$, required by the algorithm to terminate. We consider the general case where $c$ might not be strongly convex and develop an algorithm inspired from~\cite{fercoq2019adaptive} to solve the idealistic control problem with a linear rate of convergence to a solution, i.e. proportional to $\ln 1/\epsilon$. We deduce a maximum number of flops, quadratic in both $n$ and $m$ and linear in the length of the trajectory $\mathscr{T}_p$, required by \controlalgo{} to compute near-optimal control values when the idealistic formulation is used.

\algdef{SE}[DOWHILE]{Do}{doWhile}{\algorithmicdo}[1]{\algorithmicwhile\ #1}%
\begin{algorithm}[!t]
    \caption{\optsolver{}: Accelerated proximal gradient with adaptive restart for optimization problems satisfying the hypothesis of Lemma~\ref{lem:bound-general-setting}}\label{algo:adaRes}
    \begin{algorithmic}[1]    
    \Require{Desired accuracy $\epsilon$ for the optimality, a point $y_0 \in \mathbb{R}^p$, smoothness parameter $L$, an initial estimation $\mu_0$ of $\mu_{\mathcal{P}}(y_0)$, and a function $\Pi_{\mathcal{P}}$ for projection onto $\mathcal{P}$.}
    \Ensure{$\hat{y} \in \mathcal{P}$ such that $l(\hat{y}) - l^{\star} \leq \epsilon$} 
        \State $s \gets -1$, $t_s \gets 0$, $y_{s,t_s} \gets y_0$, $y_{s+1,0} \gets \Pi_{\mathcal{P}}(y_{s,t_s})$
        \Do
            \State $s \gets s+1$, $t \gets 0$
            \State $C_s \gets 16 \mu_s^{-1} \|y_{s,0} - y_{s-1, t_{s-1}}\|^2_2$
            \State $K_s \gets 2 \sqrt{e \mu_s^{-1}} -1$
            \Do
                \State $x_0 \gets y_{s,t}$, $z_0 \gets x_0$, $\theta_0 \gets 1$
                \For{$k \in \mathbb{N}_{[0,\hdots, K_s -1]}$}
                    \State $x_{k+1} \gets \Pi_{\mathcal{P}}(z_k - \frac{1}{L} \nabla l (z_k) ) $ \label{alg:apg-deb}
                    \State $\theta_{k+1}$ solves $\theta_{k+1} = (1 - \theta_{k+1}) \theta_{k}^2$
                    \State $\beta_{k+1} = \theta_k (1-\theta_k) / (\theta_k^2 + \theta_{k+1})$
                    \State $z_{k+1} = x_{k+1} + \beta_{k+1} (x_{k+1} - x_k)$\label{alg:apg-end}
                \EndFor
                \State $y_{s,t+1} \gets z_{K_s}$, $t \gets t+1$
            \doWhile{ $\|\Pi_{\mathcal{P}}(y_{s,t}) - y_{s,t} \|^2_2 > C_s (\theta^2_{K_s-1} \mu_s^{-1})^t$ or  $\qquad \qquad L^2 \|\Pi_{\mathcal{P}}(y_{s,t}) - y_{s,t}\|^2_2 \leq \epsilon \mu_s / 8$}
            \State $t_s \gets t$, $y_{s+1,0} \gets \Pi_{\mathcal{P}}(y_{s,t_s})$, $\mu_{s+1} \gets \mu_s / 2$
        \doWhile{$L^2 \|y_{s+1,0} - y_{s,t_s}\|^2_2 \leq \epsilon \mu_s / 8$}
        \State $\hat{N} \gets 1 + \sum_{k=0}^s (t_s K_s + 1)$
        \State \Return $\Pi_{\mathcal{P}}(y_{s, t_s})$
  \end{algorithmic}
\end{algorithm}

\begin{lemma}[\textsc{\cite{fercoq2019adaptive}, Theorem 2}]\label{lem:bound-general-setting}
    Consider the convex optimization problem $\underset{y \in \mathcal{P}}{\mathrm{inf}} \quad l(y)$, where $\mathcal{P} \subseteq \mathbb{R}^p$ is a convex set and $l : \mathbb{R}^p \mapsto \mathbb{R}$ is a differentiable convex function. Let $l^{*} \in \mathbb{R}$ be the optimal cost, $\mathcal{P}^{\star}$  \myet{the set of minimizers} of the problem, and assume that there exists $L > 0$ and $\mu_{\mathcal{P}}(y_0) > 0$ for any $y_0 \in \mathbb{R}^p$ such that  $l$ satisfies the smoothness and local error bound assumption given respectively by
    \begin{align}
        &l(y) \leq l(w) + \nabla l (w)^{\mathrm{T}} (y-w) + \frac{L}{2} \|y-w\|_2,\label{eq:smoothness-ass}\\
        &l(z) \geq l^{\star} + \frac{\mu_{\mathcal{P}}(y_0)}{2} \mathrm{dist}^2(z, \mathcal{P}^{\star}), \label{eq:local-quad-growth} 
    \end{align}
    for all $y, w \in \mathbb{R}^p$ and for all $z \in \mathcal{P}$ such that $l(z) \leq l(y_0)$. Then, for any $\mu_0 >0$ and $\epsilon > 0$,  Algorithm~\ref{algo:adaRes} returns a solution $\hat{y} \in \mathcal{P}$ such that  $l(\hat{y}) - l^{\star} \leq \epsilon$ with a total number of iterations $\hat{N}$ satisfying 
        \begin{align}
        \hat{N} &\leq \Big\lceil{2\sqrt{\frac{e}{\mu_0}}-1}\Big\rceil \Big\lceil \ln \frac{16(l(y_0) - l^{\star})}{\epsilon \mu_0} \Big\rceil, \: \mathrm{if} \: \:\mu_0 \leq \mu_{\mathcal{P}}(y_0), \nonumber\\
        \hat{N} &\leq 1 + \frac{2\sqrt{e}}{\sqrt{2}-1} \Big( \sqrt{\frac{2}{\mu_{\mathcal{P}}(y_0)}} - \sqrt{\frac{1}{\mu_0}} \Big) \Big\lceil \ln \frac{2^8(l(y_0) - l^{\star})}{\epsilon \mu_{\mathcal{P}}( y_0)^{2}}\Big\rceil\nonumber \\
                & \quad \quad + \Big\lceil \frac{16\sqrt{2e}}{\sqrt{\mu_{\mathcal{P}}( y_0)}}\Big\rceil\Big\lceil \ln \frac{2(l(y_0) - l^{\star}}{\epsilon \mu_{\mathcal{P}}(y_0)}\Big\rceil, \:\: \mathrm{otherwise}. \label{eq:max-iterations}
    \end{align}
\end{lemma}
\begin{remark}
    Lemma~\ref{lem:bound-general-setting} guarantees a linear rate of convergence of \optsolver{} without knowing explicitly the parameter $\mu_{\mathcal{P}}(y_0)$. Specifically, \optsolver{} achieves this rate of convergence by being able to check at iteration $s$ whether the current estimate $\mu_s$ of $\mu_{\mathcal{P}}(y_0)$ is too large and finally by refining such estimate. Therefore, \optsolver{} can also be used to estimate a lower bound on $\mu_{\mathcal{P}}(y_0)$ and thus providing the guarantees~\eqref{eq:max-iterations}. Note that the number of iterations $\hat{N}$ represents also the \myet{total number of calls} to the projection $\Pi_{\mathcal{P}}(\cdot)$ onto $\mathcal{P}$.
\end{remark}

We exploit Lemma~\ref{lem:bound-general-setting} to provide a maximum number of flops required by \controlalgo{} to terminate.
\begin{theorem}[\textsc{Number of flops by \controlalgo{}}]\label{thm:ide-bounds}
    At each sampling time $t_i \geq t_N$, given a trajectory $\mathcal{T}_i$ of length $i \geq N$, if \controlalgo{} based on the idealistic control problem~\eqref{eq:idealistic-control-problem} uses Algorithm~\ref{algo:datacontrolide} with $\mu_0 \leq \mu_{\mathcal{U}}(v_0)$ and without any side information~\ref{side:bounds-state-vectorfield}--\ref{side:partial-knowledge}, then \controlalgo{} takes at most $N^{\mathrm{ide}}$ flops to compute near-optimal control values with $N^{\mathrm{ide}}$ roughy given by
    \begin{align}
        N^{\mathrm{ide}} &= \Big\lceil{2\sqrt{\frac{e}{\mu_0}}-1}\Big\rceil \Big\lceil \ln \frac{32c_{\mathrm{max}}}{\epsilon \mu_0} \Big\rceil (2m^2 + 16m + 2) \nonumber\\
                        &\quad  + 2i (8n+9)(m+1) + 24n^2m + 8n^2 + 8m^2  \nonumber \\
                        &\quad  + 18nm + 32n + 2m + 8, \label{eq:it-bounds-ide}
    \end{align}
    where $c_{max} = |c(\mathcal{X}, \mathcal{U}, \mathcal{X})|$ is the maximum value in absolute value taken by the cost function over the range $\mathcal{X}$ and $\mathcal{U}$. Furthermore, if any of the side information~\ref{side:bounds-state-vectorfield}--\ref{side:partial-knowledge} is provided, then only a maximum of $R(n,m)$ flops will be added to $N^{\mathrm{ide}}$ where $R$ is a polynomial of maximum degree $2$.
\end{theorem}

We provide in Appendix~\ref{appendix-A} a proof of Theorem~\ref{thm:ide-bounds} where we show that~\eqref{eq:idealistic-control-problem} satisfies the assumption of Lemma~\ref{lem:bound-general-setting} \myet{in order} to bound the number of flops required by \controlalgo{}.
\begin{remark}
    Even though the optimistic control problem~\eqref{eq:optimistic-control-problem} satisfies the assumptions of Lemma~\ref{lem:bound-general-setting}, computing the projection $\Pi_{\mathcal{P}}$ is as hard as solving directly~\eqref{eq:optimistic-control-problem}. Hence, we choose to focus on the idealistic control problem. Note that $\mu_0$ can be empirically estimated offline by solving the idealistic control problem on many instances of the same problem. 
\end{remark}

\algdef{SE}[DOWHILE]{Do}{doWhile}{\algorithmicdo}[1]{\algorithmicwhile\ #1}%
\begin{algorithm}[!hbt]
    \caption{Solve the idealistic control problem}\label{algo:datacontrolide}
    \begin{algorithmic}[1]    
    \Require{Desired accuracy $\epsilon$ for optimality, a point $v_0 \in \mathbb{R}^m$, an initial estimation $\mu_0 > 0$ of $\mu_{\mathcal{U}}(v_0)$, parameters $\mathcal{B}_i$, $\mathcal{A}_i^+$, $\mathcal{A}_i^-$ from~\eqref{eq:linear-over}, weights $w^+, w^- \in [0,1]$.}
    \Ensure{$\hat{u}_i \in \mathcal{U}$ solution of~\eqref{eq:idealistic-control-problem}} 
        \State Compute $A^{\mathrm{ide}}$, $b^{ide}$ via~\eqref{eq:coeff-Ai-idealistic-pb}, $w^+$, and $w^-$
        \State Compute $Q_i$, $q_i$, $p_i$ via~\eqref{eq:Qi_ide},~\eqref{eq:qi_ide},~\eqref{eq:pi-ide}, $Q$, $R$, $S$, $q$, $r$, $A^{\mathrm{ide}}$, and $b^{\mathrm{ide}}$
        \State Define for $l \leq m$, $(\Pi_{\mathcal{U}}(v))_l = \begin{cases} (\underline{\mathcal{U}})_l, \: \: \: \text{ if } v_l \leq (\underline{\mathcal{U}})_l \\ (v)_l, \: \: \: \text{ if } (v)_l \in (\mathcal{U})_l \\ (\overline{\mathcal{U}})_l, \: \: \: \text{otherwise} \end{cases}$
        \State \Return Output of \optsolver{} with inputs $\epsilon$, $L=1/(\sqrt{m} \|Q_i\|_{\infty})$, $\Pi_{\mathcal{U}}$, and $\nabla l (z_k) = Q_i z_k + q_i$
  \end{algorithmic}
\end{algorithm}

\begin{figure*}[!t]
    \centering
    \input{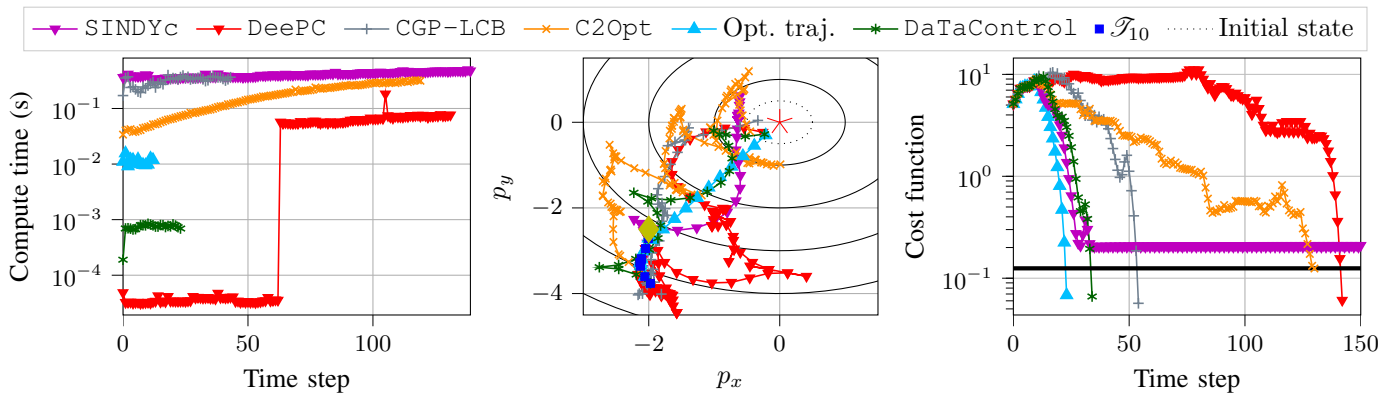}
    \vspace*{-7mm}
    \caption{\myet{\controlalgo{} outperforms state-of-the-art data-driven approaches in terms of computation and optimality of the computed control when applied to a scenario with no noise in the measurements. $\mathrm{Opt.\ traj.}$ corresponds to the one-step optimal control with the known dynamics, and the solid black line on the far right figure represents the optimality threshold for reaching the goal. We emphasize that the extremely low computation time of \texttt{DeePC} at the beginning is due to randomly generated control inputs until the satisfaction of the persistent excitation condition.}}
    \vspace*{-3mm}
    \label{fig:unicycle_traj_cost}
\end{figure*}

\section{Numerical Simulations}

We compare \controlalgo{} with existing data-driven control algorithms \texttt{CGP-LCB}~\cite{krause2011contextual,gpyopt2016}, \texttt{C2Opt}~\cite{vinod2020convexified}, \myet{ \texttt{DeePC}~\cite{Coulson2019DataEnabledPC}}, and \texttt{SINDYc}~\cite{kaiser2018sparse}. These algorithms also solve the one-step optimal control problem~\eqref{eq:one-step-optimal-control} approximately. \texttt{CGP-LCB} assumes that the unknown one-step approximate cost function $C(x_i,u_i) = c(x_i, u_i, x(t_{i+1}; x_i, u_i))$ to minimize is described by a Gaussian process,  while \texttt{C2Opt} assumes that $C$ has Lipschitz continuous gradients. On the other hand, \texttt{SINDYc} uses the limited data to perform \emph{sparse identification} of the dynamics over a library of nonlinear functions, which then permits an approximate solution to \eqref{eq:one-step-optimal-control} via numerical solvers. \myet{\texttt{DeePC} directly computes control values by solving an $N$-step model predictive control problem without the system identification step}. The experiments on the unicycle system, a quadrotor, and an aircraft system demonstrate that \controlalgo{} outperforms these algorithms both on the computation time at each sampling time\myet{, the robustness to noise, } and the suboptimality of the control. \myet{We provide the code for reproducibility of the results of this paper at the link: \href{https://github.com/wuwushrek/datacontrolreach.git}{https://github.com/wuwushrek/datacontrolreach.git}.}

\subsection{The Unicycle System}
We consider the problem of driving the unicycle of Example~\ref{ex:unicycle-system} to the origin. We encode this control objective using the one-step cost function $c(x_i,u_i,x_{i+1}) = 0.5 \|x_{i+1}\|^2_2$. We chose the time step size $\Delta t = 0.1s$ and generated a random initial trajectory $\mathscr{T}_{10}$ starting from the state $x_0= [-2,-2.5,\pi/2]$ and such that the unicycle goes away from the target. Note that the time step size $\Delta t = 0.1s$ is within the limit enforced by~\eqref{eq:inf-bounds-step-size} as $1/(\sqrt{n} \beta_{\inf}) \approx 0.123$. Hence, we used the a priori rough enclosure given by Theorem~\ref{thm:step-size-fixpoint}.

We applied \controlalgo{} with the side information discussed in Example~\ref{ex:unicycle-system}. We use \controlalgo{} with Algorithm~\ref{algo:datacontrolide} as the baseline to solve the idealistic control problem. The weights $w^+$ and $w^-$ were chosen randomly, and $\mu_0 = 1.0$. We used the default parameters in \texttt{GPyOpt}~\cite{gpyopt2016} when implementing \texttt{CGP-LCB}. We chose the Lipschitz constant for the gradient $L=10$ and trade-off hyperparameter $\alpha=1/2$ for \texttt{C2Opt}~\cite{vinod2020convexified}. For \texttt{SINDYc}, we considered monomials (up to degree $6$), sines and cosines of the state, and the products of these functions with the velocity $v$ and the turning rate $\omega$  as the library functions. To perform sparse identification, we swept the regularization parameter~\cite{kaiser2018sparse} $\lambda\in\{10^{p}:p\in \mathbb{N}_{[-6,5]}\}$ and rounded-down the coefficients smaller than $10^{-3}$ to zero. \myet{For \texttt{DeePC}, we use the data-driven model predictive control scheme $(8)$ in~\cite{Coulson2019DataEnabledPC} with the slack variable and the regularization in the cost to account for the nonlinearities in the unicycle dynamics. After extensive tuning, we use the parameters $\lambda_{y} = 10^4$, $\lambda_{g} = 1$, $T_{\mathrm{ini}} = 8$, and the horizon control $N=10$. We note that finding working hyperparameters for nonlinear dynamics was a hard task and the hyperparameters did not work for different initializations of the unicycle.} We relaxed the target state to the $0.1$-sublevel set of the one-step cost function as the stopping criteria for the algorithms.

Figure~\ref{fig:unicycle_traj_cost} shows the trajectories and the evolutions of the cost functions for the different algorithms \myet{under noiseless measurements.} \controlalgo{} performs significantly better than \texttt{GPyOpt}, \texttt{SINDYc}, \myet{\texttt{DeePC},} and \texttt{C2Opt}. It reaches the origin in fewer time steps and significantly lower computation time. Instead, \texttt{SINDYc} failed as it gets stuck close to the origin and is unable to further improve its one-step cost function.

\begin{figure}[!hbt]
    \centering
    \hspace*{-3mm}
    \input{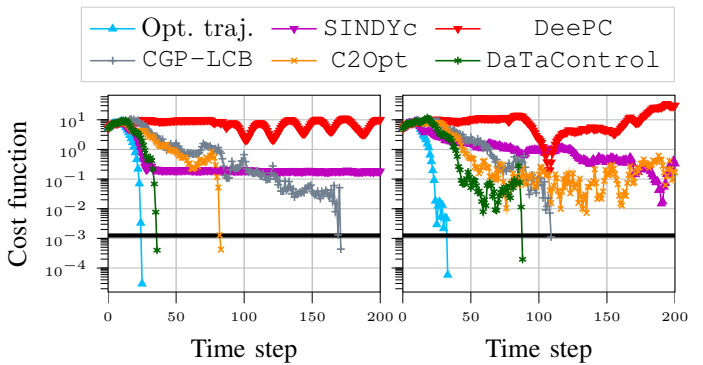}
    \vspace*{-7mm}
    \caption{\myet{\controlalgo{} outperforms the baselines in terms of optimality and robustness when applied to scenario with different levels of noise in the measurements. On the left figure, the noise on the measurements of the state and the state derivative is from a uniform distribution with bounds $\eta = [0.01,0.01,0.01]$ and $\bar{\eta} = [0.1,0.1,0.1]$, respectively. On the right figure, we have the bound on the noise $\eta = [0.1,0.1,0.05]$ and $\bar{\eta} = [1,1,0.5]$.}}
    \label{fig:unicyclenoise}
\end{figure}

\myet{Additionally, we compare \controlalgo{} with the data-driven baselines under the same conditions as the previous experiment but with different levels of noise in the measurements of the state and the state derivatives. Figure~\ref{fig:unicyclenoise} shows that \controlalgo{} has improved robustness to noise.}
\begin{figure*}[!t]
    \centering
    \input{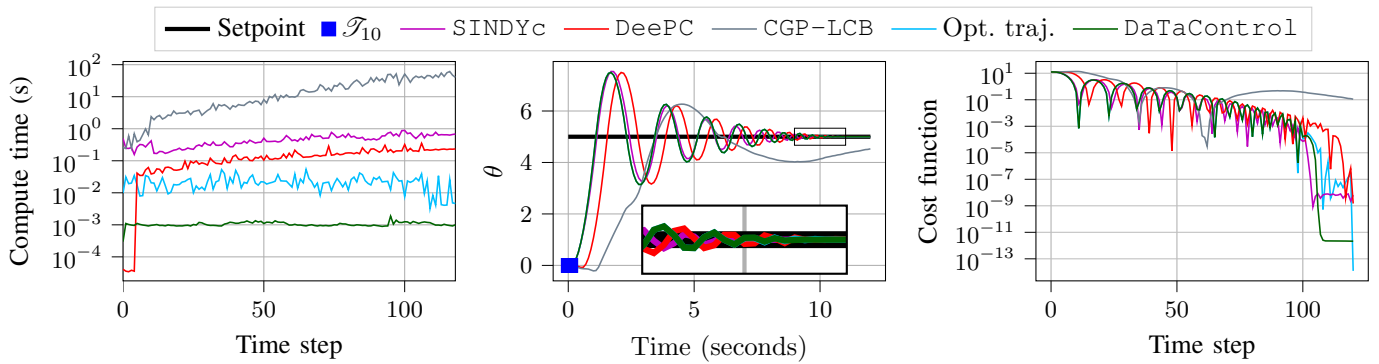}
    \vspace*{-7mm}
    \caption{\myet{\controlalgo{} outperforms the baselines in terms of computation time and optimality of the control even on a linear system in a noise-free scenario.}}
    \vspace*{-3mm}
    \label{fig:aircraftnonoise}
\end{figure*}

\subsection{Damaged Aircraft System}
In this section, we consider the problem of controlling \myet{an aircraft with nonlinearities introduced as damage in the pitch.} 
\begin{align}
        \dot{x}_1 &= -0.021 x_1 + 0.122 x_2 - 0.322 x_3 + 0.01 u_1 + u_2, \nonumber\\
        \dot{x}_2 &= -0.209 x_1 -0.53 x_2 + 2.21 x_3 - 0.064 u_1 - 0.044 u_2, \nonumber\\
        \dot{x}_3 &= 0.017 x_1 + 0.01 \cos(x_1)x_1 - 0.164 x_2 + 0.15 \sin(x_1) x_2, \nonumber\\
                 & \quad -0.421 x_3 -0.378 u_1 + 0.544 u_2 + 0.5 \sin(x_2) u_2, \nonumber\\
        \dot{x}_4 &= x_3, \quad \quad \quad  \dot{x}_5= -x_2 + 2.21 x_4.  \nonumber
\end{align}
The model above is based on the linearized dynamics of an undamaged Boeing 747 in landing configuration at sea level~\cite{bryson1993control,che2001automatic}, where the damages are introduced via nonlinear factors in the dynamics~\cite{ornik2019myopic}. The state of the aircraft is given by $x = [w_l, w_v,q,\theta, h]$ and the control input by $u = [\delta_e,\delta_t]$. $w_l$ is the deviation from the steady-state longitudinal speed, $w_v$ the downwards-pointing vertical speed, $q$ the pitch rate, $\theta$ the pitch angle and $h$ the altitude. The control inputs $\delta_e$ and $\delta_t$ are the deviations from the elevator and thrust values.

The introduced damages emulate damages to the horizontal stabilizer of the aircraft, which result in the pitch rate behaving more erratically. We seek to control the pitch angle to a setpoint $\theta^{\mathrm{sp}} = 5$ while assuming that the dynamics are unknown. We encode this control objective using the cost function $c(x_i, u_i, x_{i+1}) = 0.5 ((x_{i+1})_4 - \theta^{\mathrm{sp}})^2$. We chose the time step size $\Delta t = 0.01s$ and generated a random initial trajectory $\mathscr{T}_{10}$ starting from the state $x_0= [0,0,0,0,100.0]$. \myet{We used \controlalgo{}, \texttt{CGP-LCB} with the same hyperparameters as for the unicycle dynamics. For \texttt{SINDYc}, we considered monomials (up to degree $2$), sines and cosines of the state. We did not investigate \texttt{C2Opt} due to the inability to compute the gradient of $C$ and the fact that it is a strong assumption to assume measurements of such gradient.} For \controlalgo{}, we consider the elementary side information $\dot{\theta} = q$. Furthermore, we consider that $G(x) = G(q,w_v)$ and $f(x) = f(w_l, w_v, q, \theta)$ as the longitudinal speed and the altitude do not directly affect the derivative of the states. We considered the loose Lipschitz bounds $L_f = [0.4, 3, 4, 1, 3]$, $L_{G_{3,2}} = 0.5$, and $L_{G_{k,l}} = 0.01$ otherwise.

\myet{We first consider a linear version of the aircraft dynamics where we remove the nonlinear terms $\sin(x_1)$, $\cos(x_1)$, and $\sin(x_2)$ introduced for the damage. The objective here is to show that \controlalgo{} is more performant on linear systems than state-of-the art approaches based on behavioral systems theory foundation, e.g., \texttt{DeePC}. For linear systems, \texttt{DeePC} does not use slack variables and considers the parameters $\lambda_{y} = 0$, $\lambda_{g} = 0$ and $T_{\mathrm{ini}} = 8$ as instructed in~\cite{Coulson2019DataEnabledPC}. Figure~\ref{fig:aircraftnonoise} shows the near-optimality of \controlalgo{}, while $\texttt{CGP-LCB}$ fails to stabilize the pitch in a noise-free scenario.}

\begin{figure}[!hbt]
    \centering
    \hspace*{-3mm}
    \input{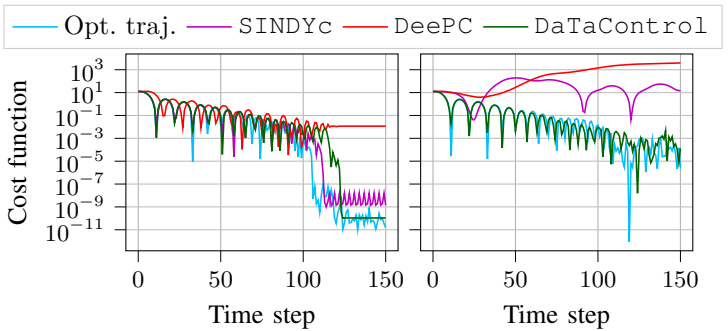}
    \vspace*{-7mm}
    \caption{\myet{\controlalgo{} outperforms the baselines in terms of optimality and robustness when applied to scenarios with different levels of noise in the measurements of the nonlinear aircraft dynamics. On the left figure, we consider a scenario with noise-free measurements. On the right figure, the noise is from a uniform distribution with the bound $\eta = \bar{\eta} = [0.01, 0.01, 0.01, 0.01, 0.1]$.}}
    \vspace*{-7mm}
    \label{fig:aircraft-ex}
\end{figure}

\myet{Then, we consider the nonlinear aircraft dynamics above with bounded noise on the measurements on the state and the state derivatives. In this scenario, we use the parameters $\lambda_{y} = 10^3$, $\lambda_{g} = 10$ to account for nonlinearities when using \texttt{DeePC}. Figure~\ref{fig:aircraft-ex} shows that \texttt{DeePC} is less performant on the nonlinear dynamics, while \texttt{SINDYc} and \texttt{DeePC} fail when there is noise in the measurements. We did not compare with \texttt{CGP-LCB} due to its extremely high computation time for solving the control problem at each time step.}
\begin{figure*}[t]
    \centering
    \input{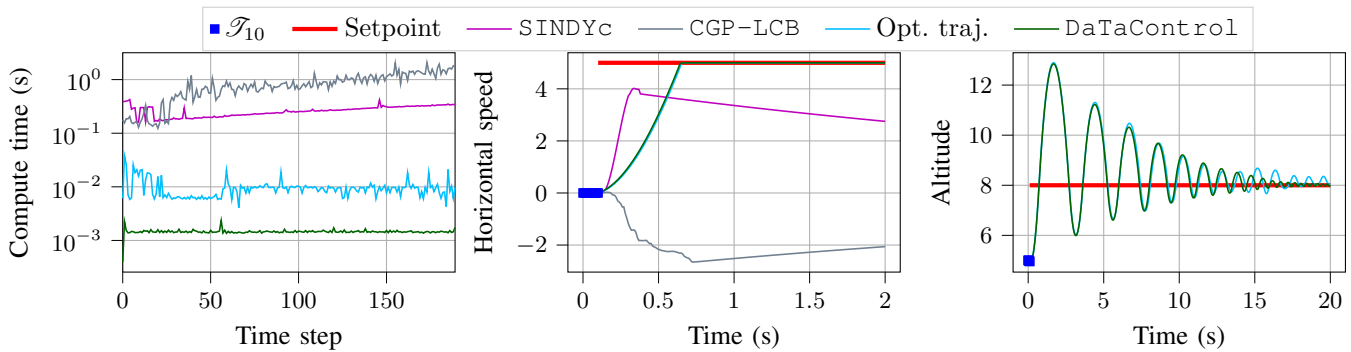}
    \vspace*{-5mm}
    \caption{\texttt{CGP-LCB} and \texttt{SINDYc} fail in the task of controlling the horizontal speed $v_x$ of the quadrotor whereas \controlalgo{} reaches both setpoints.}
    \vspace*{-3mm}
    \label{fig:quadrotor_traj_cost}
\end{figure*}

\subsection{Quadrotor System}\label{sec:quadrotor}
Consider a quadrotor with control-affine dynamics~\cite{decomposition-tomlin}
\begin{align*}
    &\dot{p}_x = v_x, &\dot{v}_x = -\frac{1}{m} C^v_D v_x-\frac{T_1}{m} sin \phi -\frac{T_2}{m} sin \phi, \\
    &\dot{p}_y = v_y, &\dot{v}_y = -\frac{1}{m}(m g+C^v_D v_y) + \frac{T_1}{m} cos \phi + \frac{T_2}{m} cos \phi, \\
    &\dot{\phi} = \omega,  &\dot{\omega} = -\frac{1}{2I_{yy}} C^\phi_D \omega -\frac{l}{2I_{yy}}T_1+\frac{l}{2I_{yy}} T_2,
\end{align*}
where the components of the state $x = [p_x,v_x,p_y,v_y,\phi,\omega]$ represent, respectively, the horizontal position, horizontal velocity, vertical position, vertical velocity, pitch angle, and pitch rate. The components of the control $u = [T_1, T_2]$ represent the thrust exerted on either end of the quadrotor. We chose the constraint set $\mathcal{U} = [0,18.4] \times [0,18.4]$.  The constants of the dynamics are given by $C^v_D=0.25$, $C^\phi_D=0.02255$, $g=9.81$, $m=1.25$, $l=0.5$, and $I_{yy}=0.03$.

We assume that the dynamics of the quadrotor are unknown and consider the problems of controlling $v_x$ to a setpoint $v_x^{\mathrm{sp}} = 5$ and $p_y$ to a setpoint $p_y^{\mathrm{sp}} = 8$. We encode these control objectives, respectively, using the cost functions $c_1(x_i,u_i,x_{i+1}) = 0.5 ((x_{i+1})_2 - v_x^{\mathrm{sp}})^2$ and $c_2(x_i,u_i,x_{i+1}) = 0.5 ((x_{i+1})_3 - p_y^{\mathrm{sp}})^2$. We chose the time step size $\Delta t = 0.01s$ and generated a random initial trajectory $\mathscr{T}_{10}$ starting from the state $x_0= [0,0,5,0,0,0]$. We consider the same hyper-parameters for Algorithm~\ref{algo:datacontrolide} as in the unicycle case. We applied \controlalgo{} with the side information $\dot{p}_x = v_x$, $\dot{p}_y = v_y$, and $\dot{\phi} = \omega$. Such extra knowledge is obtained from elementary laws of physics. Furthermore, \controlalgo{} considered the loose Lipschitz bounds $L_{f_2}= L_{f_4} = 0.3$, $L_{f_6} = 0.9$, $L_{G_{6,1}}= L_{G_{6,2}} = 0.01$, $L_{G_{2,1}}= L_{G_{2,2}} = L_{G_{4,1}}= L_{G_{4,2}} = L_{G_{6,1}}= L_{G_{6,2}} = 0.9$, and uses the side information $G(x) = G(\phi)$ and $f(x) = f(v_x,v_y,w)$. We used the default parameters of \texttt{GPyOpt} when implementing \texttt{CGP-LCB}. For \texttt{SINDYc}, we considered monomials (up to degree $1$), sines and cosines of the state, and the products of these functions with the $T_1$ and $T_2$ as the library functions. We did not investigate \texttt{C2Opt} due to the inability to compute the gradient of $C$.

Figure~\ref{fig:quadrotor_traj_cost} shows the near-optimality of \controlalgo{} while \texttt{CGP-LCB} and \texttt{SINDYc} fail to reach the setpoints. Furthermore, the figure demonstrates that \controlalgo{} can achieve near-optimal, near-real-time control of the vertical position and horizontal speed. We justify the suboptimality of $\mathrm{Opt.\ traj.}$ by the fact that the one-step optimal control problem is highly nonlinear, which makes possible the synthesis of local optimum solutions by the numerical solvers.

\section{Conclusion}
We developed two data-driven algorithms, \reachalgo{} and \controlalgo{}, for on-the-fly over-approximation of the reachable set and constrained near-optimal, real-time control of systems with unknown dynamics. These algorithms are suitable for scenarios with severely limited \myet{noisy} data and can take advantage of various side information on the underlying a priori unknown dynamics. The numerical experiments demonstrate the efficacy of the algorithms over existing approaches both in terms of the suboptimality of the control, \myet{robustness to noise,} and the computation time to synthesize control values.

\bibliographystyle{IEEEtran}
\bibliography{IEEEabrv,ref}

\begin{appendices}

\section{}\label{appendix-A}
In this appendix, we provide the proofs for all lemmas and theorems presented in this paper.

\section*{Proof of Theorem~\ref{thm:step-size-fixpoint}}

First, we provide a bound on the maximum variation of the function $h$ given upper bounds on the Lipschitz constants of $f$ and $G$.

\begin{lemma}[\textsc{Maximum variation of $h$}] \label{lem:h-variation}
Let $L_{f}$ and $L_{G}$ be upper bounds on the Lipschitz constants of $f$ and $G$. For all $x_1,x_2 \in \mathcal{X}$ and $u_1 \in \mathcal{U}$, we have that
\begin{align} \label{eq:maximum-variation-h}
    \| h(x_1, u_1) - h(x_2, u_1) \|_2 \leq \beta(u_1) \| x_1 - x_2 \|_2 ,
\end{align}
where $\beta(u_1) = \sqrt{ \sum_{k=1}^n (L_{f_k} + \sum_{l=1}^m L_{G_{k,l}} |{(u_1)}_l|)^2 }$.
\end{lemma}
\begin{proof}
    This is a direct application of the definition of the Lipschitz bounds of $f_k$ and $G_{k,l}$. Specifically, we have that
    \begin{align}
        &|{(h(x_1,u_1))}_k - {(h(x_2,u_1))}_k| \nonumber\\ 
        &= |{(f(x_1) - f(x_2))}_k+ \sum_{l=1}^m  {((G(x_1) - G(x_2))}_{k,l} {(u_1)}_l | \nonumber\\
        &\leq \big(L_{f_k} + \sum_{l=1}^m L_{G_{k,l}} |{(u_1)}_l| \big) \|x_1-x_2\|_2, \label{eq:bounds-hi}
    \end{align}
    for all $k \in \mathbb{N}_{[1,n]}$. The last inequality is obtained by the definitions of $L_{f_k}$ and $L_{G_{k,l}}$. Hence, we obtain ~\eqref{eq:maximum-variation-h}.
\end{proof} 

\begin{lemma}[\textsc{Grönwall's inequality~\cite[Lemma 2.7]{teschl2012ordinary}}] \label{lem:gronwall-lemma}
Let $\Psi, \alpha, \gamma$ be real-valued functions on $[t_0, T]$ with $t_0,T \in \mathbb{R}$. Suppose $\Psi$ satisfies for all $t \in [t_0,T]$ the inequality
\begin{align}
    \Psi (t) \leq \alpha (t) + \int_{t_0}^t \gamma (s) \Psi (s) ds, \label{eq:cond-gronwall-inequality}
\end{align}
with $\gamma (s) \geq 0 $ for all $s \in [t_0,T]$. Then, for all $t \in [t_0,T]$,
\begin{align} \label{eq:gronwall-inequality}
    \Psi (t) \leq \alpha (t) + \int_{t_0}^{t} \alpha (s) \gamma (s) \exp \Big( \int_{s}^t \gamma (r) dr \Big) ds.
\end{align}
\end{lemma}

Next, we combine the results of Lemma~\ref{lem:h-variation} and Lemma~\ref{lem:gronwall-lemma} to provide a bound on the variations of points in a trajectory.

\begin{lemma}[\textsc{Variation of trajectories of~\eqref{eq:control-linearization}}]\label{lem:rate-change-solution}
Let $x$ be a continuous-time signal satisfying~\eqref{eq:control-linearization} for a given control signal $u \in \mathbb{V}$. For all $t \in [t_i, t_i + \Delta t]$, we have that 
\begin{align}
    \| x(t) - x(t_i)\|_2 \leq \|\alpha_i\|_2 \beta_i^{-1} (\mathrm{e}^{\beta_i \Delta t} - 1), \label{eq:max-variation-solution}
\end{align}
where the parameters $\beta_{i} \in \mathbb{R}_+$ and $\alpha_{i} \in \mathbb{R}^n$ are given by
\begin{align}
    \beta_{i} &= \sqrt{ \sum_{k=1}^n \Big( L_{f_k} + \sum_{l=1}^m L_{G_{k,l}} |(\boldsymbol{v}([t_i,t_i+\Delta t]))_l| \Big)^2 }, \label{eq:beta-formula}\\
    \alpha_{i} &= f(x(t_i)) + |G(x(t_i))| \: |\boldsymbol{v}([t_i,t_i+\Delta t])|. \label{eq:alpha-formula}
\end{align}
\end{lemma}

\begin{proof}
    For all $t \in \mathcal{T}_i = [t_i, t_i + \Delta t]$, we have that
    \begin{align}
        &\|x(t) - x(t_i)\|_2 \nonumber \\
        &= \|\int_{t_i}^{t} h(x(s), u(s)) ds \|_2 \label{eq:pvarsol-solution-characteristic} \\
        &\leq \begin{aligned}[t]
                    &\int_{t_i}^{t} \|h(x(s),u(s))-h(x(t_i),u(s))\|_2 ds \\
                    &\quad +  \int_{t_i}^{t} \|h(x(t_i),u(s))\|_2 ds
          \end{aligned} \label{eq:pvarsol-triangle-inequality}\\
        &\leq \begin{aligned}[t]
                &\int_{t_i}^{t} \beta(u(s)) \|x(s)-x(t_i)\|_2 ds + \int_{t_i}^{t} \|h(x(t_i),u(s))\|_2 ds
            \end{aligned} \label{eq:pvarsol-lipschitz-bounds} \\
        &\leq \|\alpha_i\|_2 (t-t_i) + \int_{t_i}^t \beta_i \|x(s)-x(t_i)\|_2 ds. \label{eq:pvarsol-pre-gronwall} 
    \end{align}
    We obtain~\eqref{eq:pvarsol-solution-characteristic} since $x$ is a solution of~\eqref{eq:control-linearization}. The passage from~\eqref{eq:pvarsol-solution-characteristic} to~\eqref{eq:pvarsol-triangle-inequality} results from applying the triangle inequality. We use Lemma~\ref{lem:h-variation} with $x_1 = x(s)$, $x_2=x(t_i)$, and $u_1 = u(s)$ to obtain~\eqref{eq:pvarsol-lipschitz-bounds}. We obtain~\eqref{eq:pvarsol-pre-gronwall} by showing that the inequalities $\beta(u(s)) \leq \beta_i$ and $\|h(x(t_i),u(s))\|_2 \leq \|\alpha_i\|_2$ hold for all $s\in\mathcal{T}_i$. Specifically, the former inequality holds by upper-bounding $|{(u(s))}_l|$ in $\beta(u(s))$ with $|(\boldsymbol{v}([t_i,t_i+\Delta t]))_l| \geq \sup_{s\in\mathcal{T}_i} |{(u(s))}_l|$. The latter inequality holds since
    \begin{align*}
        &|{(h(x(t_i),u(s)))}_k| \\
        &\leq \big| {(f(x(t_i)))}_k + \sum_{l=1}^m |{(G(x(t_i))}_{k,l}| \sup_{s\in \mathcal{T}_i} |{(u(s))}_l| \big| \leq |(\alpha_i)_k|.
    \end{align*}
    The inequality~\eqref{eq:pvarsol-pre-gronwall} satisfies the condition~\eqref{eq:gronwall-inequality} of Lemma~\ref{lem:gronwall-lemma} with $\Psi (t) = \|x(t) - x(t_i)\|_2$, $\alpha (t) = \|\alpha_i\|_2 (t-t_i)$, and $\gamma(t) = \beta_i$. Thus, we have that
    \begin{align}
        \|x(t) - x(t_i)\| \leq \begin{aligned}[t]
                                    &\|\alpha_i\|_2 (t-t_i) \\
                                    &+ \beta_i \|\alpha_i\|_2 \int_{t_i}^t (s-t_i) \mathrm{e}^{\beta_i(t-s)} ds.
                                \end{aligned} \label{eq:pvarsol-post-gronwall}
    \end{align}
    By integration by parts, we have that
    \begin{align}
        \int_{t_i}^t (s-t_i) \mathrm{e}^{\beta_i(t-s)} ds &= - \frac{t-t_i}{\beta_i} + \frac{1}{\beta_i}\int_{t_i}^t \mathrm{e}^{\beta_i(t-s)} ds \nonumber\\
        &= -\frac{t-t_i}{\beta_i} + \frac{\mathrm{e}^{\beta_i (t-t_i)} - 1}{\beta_i^2}. \label{eq:pvarsol-integration-part}
    \end{align}
    Finally, by combining~\eqref{eq:pvarsol-integration-part} and~\eqref{eq:pvarsol-post-gronwall}, we obtain~\eqref{eq:max-variation-solution}.
\end{proof} 

Finally, we provide the proof of Theorem~\ref{thm:step-size-fixpoint} below.

    The expression~\eqref{eq:explict-fixpoint} is derived by scaling adequately the bound~\eqref{eq:max-variation-solution} from Lemma~\ref{lem:rate-change-solution}. Specifically, we seek for an a priori rough enclosure $\mathcal{S}_i$ such that 
    \begin{equation}\label{eq:fixpoint-to-find}
        \mathcal{S}_i = \mathcal{R}_i^+ + \mu \frac{\overline{\alpha}_i}{\beta_i} (\mathrm{e}^{\beta_i \Delta t} - 1) [-1,1]^n,
    \end{equation}
    where $\mu > 0$ is a parameter to find in order for $\mathcal{S}_i$ to satisfy the fixed-point equation~\eqref{eq:apriori-enclosure-control-affine} and $\overline{\alpha}_i$ is given by 
    $$\overline{\alpha}_i = \sup_{x(t_i) \in \mathcal{R}_i^+} \| \ f(x(t_i)) + |G(x(t_i))|  \ |\boldsymbol{v}([t_i,t_i+\Delta t])| \ \|_2.$$
    We over-approximate the set $\mathscr{R}(h, \mathcal{S}_i \times \boldsymbol{v}([t_i,t_i+\Delta t]))$ as a function of $\Delta t$, $\beta_{t_i,t_i+\Delta t}$, and $\boldsymbol{h} = \boldsymbol{f} + \boldsymbol{G} u$. For all $x_i \in \mathcal{R}_i^+$, $u_i \in \boldsymbol{v}([t_i,t_i+ \Delta t])$, and $s_i \in \mathcal{S}_i$, we have that
    \begin{align}
        \|h(s_i,u_i) - h(x_i,u_i)\|&\leq \beta_i \|s_i - x_i\| \label{eq:pfixpoint-var-h}\\
        &\leq  \sqrt{n} \mu \overline{\alpha}_i (\mathrm{e}^{\beta_i \Delta t} - 1). \label{eq:pfixpoint-bound-h}
    \end{align}
    The inequality~\eqref{eq:pfixpoint-var-h} comes from Lemma~\ref{lem:h-variation} with $\beta(u_i)$ upper-bounded by $\beta_i$ and the definition of $\mathcal{S}_i$~\eqref{eq:fixpoint-to-find} provides an upper bound on $\|s_i - x_i\|$ that yields~\eqref{eq:pfixpoint-bound-h}. Additionally, the inequality~\eqref{eq:pfixpoint-bound-h} implies that
    \begin{equation*}
    \begin{aligned}[t]
         &\mathscr{R}(h, \mathcal{S}_i \times \boldsymbol{v}([t_i,t_i+\Delta t]))\\
         &\subseteq \boldsymbol{h}(\mathcal{R}_i^+,\boldsymbol{v}([t_i,t_i+\Delta t])) + \sqrt{n} \mu \overline{\alpha}_i (\mathrm{e}^{\beta_i \Delta t} - 1) [-1,1]^n.
    \end{aligned}
    \end{equation*}
    Hence, $\mathcal{S}_i$ from~\eqref{eq:fixpoint-to-find} solves the fixed-point equation~\eqref{eq:apriori-enclosure-control-affine} if
    \begin{equation}
        \begin{aligned}[t]
            & [0, \Delta t] \Big(  \boldsymbol{h}(\mathcal{R}_i^+,\boldsymbol{v}([t_i,t_i+\Delta t])) + \sqrt{n} \mu \overline{\alpha}_i (\mathrm{e}^{\beta_i \Delta t} - 1) [-1,1]^n \Big) \\
            & \subseteq \mu \frac{\overline{\alpha}_i}{\beta_i} (\mathrm{e}^{\beta_i \Delta t} - 1) [-1,1]^n. 
        \end{aligned}\label{eq:pfixpoint-rincl}
    \end{equation}
    For notation brevity, let $c_1 = \overline{\alpha}_i (\mathrm{e}^{\beta_i \Delta t} - 1)$. Observe that $[0,\Delta t][\underline{\mathcal{A}},\overline{\mathcal{A}}] = [\min (0,\underline{\mathcal{A}}), \max (0,\overline{\mathcal{A}})]$. We use the observation to find $\mu >0$ such that the inclusion~\eqref{eq:pfixpoint-rincl} holds. That is, the inequalities 
    \begin{align*}
        &\Delta t \Big( \overline{{(\boldsymbol{h}(\mathcal{R}_i^+,\boldsymbol{v}([t_i,t_i+\Delta t])))}_k} + \sqrt{n} \mu c_1 \Big) \leq \frac{\mu c_1}{\beta_i} \\
        &\Longleftrightarrow (\frac{1}{\Delta t \beta_i} - \sqrt{n}) \mu \geq \frac{1}{c_1} \overline{{(\boldsymbol{h}(\mathcal{R}_i^+,\boldsymbol{v}([t_i,t_i+\Delta t])))}_k}
    \end{align*}
    and 
    \begin{align*}
        &\Delta t \Big( \underline{{(\boldsymbol{h}(\mathcal{R}_i^+,\boldsymbol{v}([t_i,t_i+\Delta t])))}_k} - \sqrt{n} \mu c_1 \Big) \geq -\frac{\mu c_1}{\beta_i} \\
        &\Longleftrightarrow (\sqrt{n} - \frac{1}{\Delta t \beta_i}) \mu \geq \frac{1}{c_1} \underline{{(\boldsymbol{h}(\mathcal{R}_i^+,\boldsymbol{v}([t_i,t_i+\Delta t])))}_k}
    \end{align*}
    hold for all $k \in \mathbb{N}_{[1,n]}$. Therefore, for a step size $\Delta t$ satisfying~\eqref{eq:step-size-bound}, $\mu$ given by 
    \begin{align*}
        \mu = \frac{\|\boldsymbol{f}(\mathcal{R}_i^+) + \boldsymbol{G}(\mathcal{R}_i^+) \boldsymbol{v}([t_i,t_i+\Delta t])\|_{\infty}}{c_1 (\frac{1}{\Delta t \beta_i} - \sqrt{n})}
    \end{align*}
    satisfies the above inequalities. Thus, the inclusion~\eqref{eq:pfixpoint-rincl} holds and the set $\mathcal{S}_i$ is solution of the fixed-point equation~\eqref{eq:apriori-enclosure-control-affine}. By replacing $\mu$ in $\mathcal{S}_i$ given by~\eqref{eq:fixpoint-to-find}, we obtain~\eqref{eq:explict-fixpoint}.
    
\myqedblock

\section*{Proof of Lemma~\ref{lem:constrained-qp}}
    We use the interval arithmetic to characterize exactly the belonging relations $x_{i+1} \in \mathcal{B}_i + \mathcal{A}^+_i u_i$ and $x_{i+1} \in \mathcal{B}_i + \mathcal{A}^-_i u_i$ in term of linear constraints. Specifically, since $(\mathcal{A}^+_i u_i)_k = \sum_{l=1}^m (\mathcal{A}^+_i)_{k,l} (u_i)_l$, interval arithmetic provides upper and lower bound on each term $(\mathcal{A}^+_i)_{k,l} (u_i)_l$ as follows
    \begin{equation*}
        \begin{cases}   
            (\mathcal{A}^+_i)_{k,l} (u_i)_l = [\underline{(\mathcal{A}^+_i)_{k,l}} (u_i)_l,\overline{(\mathcal{A}^+_i)_{k,l}} (u_t)_l], &\text{if }  (\mathcal{U})_l \geq 0 \\[3pt]
            
            (\mathcal{A}^+_i)_{k,l} (u_i)_l = [\overline{(\mathcal{A}^+_i)_{k,l}} (u_t)_l, \underline{(\mathcal{A}^+_i)_{k,l}} (u_i)_l], &\text{otherwise.} \end{cases}
    \end{equation*}
    We can deduce that $\overline{\mathcal{A}^+_i u_i} = A_i^{\mathrm{s}+} u_i$ and $\underline{\mathcal{A}^+_i u_i} = A_i^{\mathrm{l}+} u_i$. Similarly, it is easy to prove that $\overline{\mathcal{A}^-_i u_i} = A_i^{\mathrm{s}-} u_i$ and $\underline{\mathcal{A}^-_i u_i} = A_i^{\mathrm{l}-} u_i$. As a consequence, the optimistic control problem~\eqref{eq:optimistic-control-problem} can be immediately reformulated as the convex quadratic programming problem~\eqref{eq:optimistic-expr-qp}. 
    
    Additionally, we have that
    \begin{align*}
        &c(x_i, u_i, b_i^{\mathrm{ide}} + A_i^{\mathrm{ide}} u_i) \\
        &= u_i^{\mathrm{T}}\Big((A_i^{\mathrm{ide}})^{\mathrm{T}} Q A_i^{\mathrm{ide}} + 2 (A_i^{\mathrm{ide}})^{\mathrm{T}} S + R \Big) u_i \\
        &\quad \: + \Big(  2(b_i^{\mathrm{ide}})^{\mathrm{T}} \big( S + Q A_i^{\mathrm{ide}} \big) + q^{\mathrm{T}} A_i^{\mathrm{ide}} + r^{\mathrm{T}} \Big) u_i \\
        &\quad \: + (b_i^{\mathrm{ide}})^{\mathrm{T}} Q b_i^{\mathrm{ide}} + q^{\mathrm{T}} b_i^{\mathrm{ide}} \\
        &= 0.5 u_i^{\mathrm{T}} Q_i u_i + q_i^{\mathrm{T}} u_i + p_i.
    \end{align*}
    Hence we obtain the cost function of the idealistic control problem~\eqref{eq:qp-midpoint}. Therefore, it remains to prove that the matrix $Q_i$ is positive semidefinite. Let prove that $u_i^{\mathrm{T}} Q_i u_i \geq 0$ for all $u_i \in \mathbb{R}^m$. Since by assumption, the cost function is convex, then the matrix implied in the quadratic term is positive semidefinite. Hence, we can deduce that 
    $$ \begin{bmatrix} A_i^{\mathrm{ide}} u_i \\ u_i\end{bmatrix}^{\mathrm{T}} \begin{bmatrix} Q & S \\ S^{\mathrm{T}}  & R\end{bmatrix} \begin{bmatrix} A_i^{\mathrm{ide}} u_i \\ u_t\end{bmatrix} \geq 0, $$
    for all $u_i \in \mathbb{R}^m$. Expanding the expression above immediately provides that $u_i^{\mathrm{T}} Q_i u_i \geq 0$ for all $u_i \in \mathbb{R}^m$.
\myqedblock

\section*{Proof of Theorem~\ref{thm:ide-bounds}}
    First, using the notation of Lemma~\ref{lem:bound-general-setting}, we have that $l(y) = 0.5 y^{\mathrm{T}} Q_i y + q_i^{\mathrm{T}} y + p_i$ and $\mathcal{P} = \mathcal{U}$ for the idealistic control problem. Thus, $l$ satisfies condition~\eqref{eq:smoothness-ass} with any parameter $L \geq \lambda_{\mathrm{max}}(Q_i)$ where $\lambda_{\mathrm{max}}(Q_i)$ is the maximum eigen value of $Q_i$. Recall that since we have 
    \begin{align*}
        \| \nabla l(x) - \nabla l (y) \|^2_2 & = \| Q_i x - Q_i y\|^2_2 = (x-y)^{\mathrm{T}} Q_i^TQ_i (x-y) \\
                                          & \leq \lambda_{\mathrm{max}}^2(Q_i) \|x-y\|^2_2,
    \end{align*}
    then Lemma 3.4 in~\cite{bubeck2014convex} implies that the condition~\eqref{eq:smoothness-ass} holds. Since $\lambda_{\max}(Q_i) \leq \sqrt{m} \|Q_i\|_\infty$, we pick $L = \sqrt{m} \|Q_i\|_\infty$. The local error bound assumption holds due to Theorem 2.1 in~\cite{luo1992linear} where the authors prove that for a function $h$ satisfying $h(x) = g(Ax) + c^{\mathrm{T}} x$ on a convex polyhedra $\mathcal{P}$ with $g$ strongly convex, there exists $\mu_{P}(x_0)$ for any $x_0$ such that the condition~\eqref{eq:local-quad-growth} holds. As a consequence, since we assume that $\mu_0 \leq \mu_{P}(x_0)$, Lemma~\ref{lem:bound-general-setting} provides that the number of iterations $\hat{N}$ required by \optsolver{} is upper bounded as follows
    \begin{align*}
        \hat{N} &\leq \Big\lceil{2\sqrt{\frac{e}{\mu_0}}-1}\Big\rceil \Big\lceil \ln \frac{16 (2c_{\mathrm{max}})}{\epsilon \mu_0} \Big\rceil = N_{\mathrm{max}}.
    \end{align*}
    As a consequence, since the computation of $\Pi_{\mathcal{P}}(\cdot)$ is simple and requires at most $3m$ comparisons, and $\nabla l$ requires $2m^2 + m$ flops,  we can deduce that the maximum number of flops required by \optsolver{} to terminate is roughly given by $N_{\mathrm{max}} (2m^2 + 16m + 12)$. Such a number is obtained by considering the worst case where each line of Algorithm~\ref{lem:bound-general-setting} is executed $N_{\mathrm{max}}$ times even though in reality, only lines~\ref{alg:apg-deb}--\ref{alg:apg-end} of \optsolver{} will be executed maximum $N_{\mathrm{max}}$ times. 
    
    Finally, since computing $\boldsymbol{f}$ and $\boldsymbol{G}$ requires $N(8n+9)$ and $N(8n+9)m$ respectively, we can deduce that, without any side information~\ref{side:bounds-state-vectorfield}--\ref{side:partial-knowledge}, lines~\ref{alg:datacontrol-deb}--\ref{alg:datacontrol-end}  of \controlalgo{} require at most $2N(8n+9)(m+1)+ 16n^2m + 8n^2 +2m^2 + 6nm +24n+8$. Specifically, it is easy to check that $\mathcal{S}_i$ using Theorem~\ref{thm:step-size-fixpoint} can be computed using $6mn + 6n + 4$ flops. The matrices $\mathcal{B}_i$, $\mathcal{A}^+_i$ and $\mathcal{A}^-_i$ can be computed using $4n^2 + 6n$ flops for $\mathcal{B}_i$ and $8n^2m+3nm + m^2$ flops for each $\mathcal{A}^+_i$ and $\mathcal{A}^-_i$. Hence, since computing $\mathcal{A}^{ide}_i$ and $\mathcal{B}^{\mathrm{ide}}$ requires $8n + 8nm$ flops, we can compute $Q_i$ and $q_i$ using $6n^2m + 6m^2$ and $(2n^2+4n+2)m$ flops, respectively. Thus, after adding the number of flops for \optsolver{}, lines~\ref{alg:datacontrol-deb}--\ref{alg:datacontrol-end}  of \controlalgo{}, $\mathcal{A}^{ide}_i$ and $\mathcal{B}^{\mathrm{ide}}$, and $Q_i$ and $q_i$, we obtain the bound $N^{\mathrm{ide}}$.
\myqedblock

\section*{Illustration for Lemma~\ref{lem:contraction}}
    Consider the unicycle system of Example~\ref{ex:unicycle-system}, and assume that $(G)_{1,2}$ is unknown. Given the data point $(x_i, \dot{x}_i, u_i)$ where $x_i = [0,0, \pi/2]$, $u_i = [1, 0.1]$  and $\dot{x}_i = [0, 1, 0.1] $, and the over-approximating ranges $(\mathcal{F}_i)_1 = [-0.01,1]$, $(\mathcal{G}_i)_{1,1} = [-0.05,0.05]$, and $(\mathcal{G}_i)_{1,2}=[-0.1,1]$ of $(f(x_i))_1 = 0$, $(G(x_i))_{1,1} = 0$, and $(G(x_i))_{1,2}=0$. By Lemma~\ref{lem:contraction}, we have
    \begin{align*}
        (C_{\mathcal{F}_i})_1 &= [-0.01,1] \cap (0 - [-0.05,0.05] - 0.1 [-0.1,1]) \\
        &= [-0.01,1] \cap [-0.15, 0.06] = [-0.01, 0.06].
    \end{align*}
    Hence, $(s_0)_1 = [-0.06, 0.01] \cap [-0.06,0.15] = [-0.06, 0.01] $ and the interval $(C_{\mathcal{G}_i})_{1,1}$ is given by
    \begin{align*}
        (C_{\mathcal{G}_i})_{1,1} &= \big(([-0.06,0.01] - 0.1[-0.1,1]) \cap [-0.05,0.05]\big) \\
        &= [-0.16, 0.02] \cap [-0.05,0.05] = [-0.05, 0.02].
    \end{align*}
    Moreover, since
    \begin{align*}
        (s_1)_1 &= ([-0.06,0.01] - [-0.05,0.02]) \cap (0.1 [-0.1,1] ) \\
                &= [-0.08,0.06] \cap [-0.01, 0.1] = [-0.01, 0.06],
    \end{align*} 
    the interval $(C_{\mathcal{G}_i})_{1,2}$ is given by
    \begin{align*}
        (C_{\mathcal{G}_i})_{1,2} &= 10 \big( [-0.01,0.06] \cap (0.1 [-0.1,1]) \big)= [-0.1,0.6].
    \end{align*}
    Therefore, Lemma~\ref{lem:contraction} provides the contractions $(C_{\mathcal{F}})_1 \subset (\mathcal{F}_i)_1$, $(C_{\mathcal{G}_i})_{1,1} \subset (\mathcal{G}_i)_{1,1}$, and $(C_{\mathcal{G}_i})_{1,2} \subset (\mathcal{G}_i)_{1,2}$.\vspace*{-8mm}
\myqedblock

\end{appendices}        

\end{document}